%% file: Article.tex
\documentclass[prd,amsmath,amssymb,nofootinbib,superscriptaddress,showkeys,reprint]{revtex4-2}
\usepackage{graphicx}
\usepackage{booktabs}
\usepackage{array}
\usepackage{multirow}
\usepackage{dcolumn}
\usepackage{bm}
\usepackage{cancel}

\usepackage[caption=false]{subfig}
\usepackage[table]{xcolor}
\definecolor{MAGENTA}{named}{magenta}
\usepackage[colorlinks=true, linkcolor=blue, citecolor=blue, urlcolor=blue]{hyperref}
\usepackage{orcidlink}
\usepackage{siunitx}
\newcommand{\dd}{\mathrm{d}}

\newcolumntype{B}{>{\bfseries}c}

\begin{document}

\title{Thermodynamics and $P$-$V$ Criticality of Charged AdS Black Holes with a Cloud of Strings in Kalb-Ramond Gravity}

\author{Faizuddin Ahmed\orcidlink{0000-0003-2196-9622}} 
\email[Faizuddin Ahmed - ]{faizuddinahmed15@gmail.com} 
\affiliation{Department of Physics, The Assam Royal Global University, Guwahati, 781035, Assam, India}

\author{Edilberto O. Silva\orcidlink{0000-0002-0297-5747}}
\email[Edilberto O. Silva - ]{edilberto.silva@ufma.br (Corresp. author)}
\affiliation{Programa de P\'os-Gradua\c c\~ao em F\'{\i}sica \& Coordena\c c\~ao do Curso de F\'{\i}sica -- Bacharelado, Universidade Federal do Maranh\~{a}o, 65085-580 S\~{a}o Lu\'{\i}s, Maranh\~{a}o, Brazil}

\date{\today}

\begin{abstract}
We investigate the extended phase-space thermodynamics and $P$--$V$ criticality of electrically charged anti-de Sitter (AdS) black holes in Kalb--Ramond bumblebee gravity in the presence of a spherically symmetric cloud of strings. The background Kalb--Ramond field induces Lorentz symmetry violation through a dimensionless parameter $\ell$, while the string cloud is characterized by a parameter $\alpha$, both entering the lapse function and deforming the geometry. Interpreting the ADM mass as enthalpy, we derive the main thermodynamic quantities, Hawking temperature, entropy, thermodynamic volume, Gibbs free energy, internal energy, and specific heat, and analyze how $(\ell,\alpha)$ jointly affect stability and phase structure. We obtain a Van der Waals--type equation of state, compute the critical point $(P_c,T_c,v_c)$, and show that, although the critical scales depend nontrivially on $\ell$ and $\alpha$, the universal ratio $P_c v_c/T_c = 3/8$ is preserved. Using the thermodynamic topology approach, we determine that the total topological charge of the black hole solution remains $W=1$, placing it in the same topological class as the Reissner--Nordström--AdS black hole. Finally, by studying the Joule--Thomson expansion, we derive the inversion curve and find that the minimal inversion temperature satisfies $T_i^{\rm min}/T_c = 1/2$, as in the Reissner--Nordström--AdS case, indicating that Lorentz violation and the string cloud deform the thermodynamic scales without changing the underlying universality class.
\end{abstract}

\maketitle

\section{Introduction}\label{introduction}

Black hole (BH) thermodynamics provides a powerful bridge between gravitation, statistical physics, and quantum theory. The discovery that BHs radiate with a temperature proportional to their surface gravity and possess an entropy proportional to the horizon area \cite{Bekenstein1973,Bardeen1973,Hawking1975} established that gravitational configurations can undergo phase transitions and exhibit rich thermodynamic behavior. In asymptotically anti--de Sitter (AdS) space, the interplay between the negative cosmological constant and the horizon geometry leads to phenomena such as the Hawking--Page transition between thermal AdS and large BHs \cite{HawkingPage1983}, and to an analogy between charged AdS BHs and Van der Waals fluids when the cosmological constant is promoted to a thermodynamic pressure in the extended phase space \cite{Kastor2009,Dolan2011,KubiznakMann2012,Gunasekaran2012,KubiznakMannTeo2017,Chamblin1999a,Chamblin1999b,Caldarelli2000,Cai2013GB,Wei2013GB,Hendi2015,Hendi2017,Frassino2014,Dolan2014,Hennigar2017,Altamirano2013,Altamirano2014}. In this framework, the BH mass is interpreted as enthalpy, and one finds nontrivial $P$--$V$ criticality, critical exponents of mean--field type, and first--order small--/large--BH transitions encoded in swallow--tail structures of the Gibbs free energy.

More recently, the extended phase space approach has been applied to a wide class of modified BH solutions, including those arising in alternative theories of gravity and in the presence of nontrivial matter sources (see, e.g., \cite{Cai2013,Cai2013GB,Wei2013GB,Frassino2014,Hendi2015,Hendi2017} for Gauss--Bonnet gravity, Lovelock models, and nonlinear electrodynamics). In particular, the Joule-Thomson (JT) expansion of AdS BHs has attracted considerable interest, allowing one to define inversion curves, minimum inversion temperatures, and to compare the cooling/heating behavior of BHs with that of real gases \cite{PhysRevD.102.044028,Okcu2017,Mo2018,Cisterna2018,Yekta2019,Mirza2022,Okcu2017,Okcu2018,Mo2018PRD,Lan2018PRD,MoLi2020}. At the same time, a complementary perspective based on thermodynamic topology has been developed, in which BH phases are interpreted as topological defects in a suitable thermodynamic parameter space, characterized by winding numbers and topological charges associated with generalized free energies \cite{Wei2022a,Wei2022b,Bai2023,Rizwan2023,Yerra2022}. This viewpoint offers a global, coordinate--independent classification of phase structures, beyond the local analysis of specific heats or equations of state.

On another front, Lorentz--violating (LV) extensions of gravity have been extensively investigated as low--energy manifestations of more fundamental high--energy theories. In the so--called bumblebee models, a vector or tensor field acquires a nonzero vacuum expectation value (VEV), spontaneously breaking local Lorentz symmetry and modifying the gravitational dynamics \cite{act1,act2,Kostelecky2004,Bluhm2008,Seifert2010,Casana2018,MalufNeves2021}. When the LV sector is realized by an antisymmetric Kalb--Ramond (KR) field, the corresponding Einstein--Kalb--Ramond (EKR) gravity leads to BH solutions whose metric functions and effective couplings depend explicitly on LV parameters. In Ref.~\cite{ref1}, Yang \textit{et al.} constructed neutral, static, spherically symmetric BH solutions in such a KR--induced LV theory, both with and without a cosmological constant. In Ref.~\cite{ref2}, Duan \textit{et al.} extended this analysis to electrically charged AdS BHs, deriving exact solutions and exploring their thermodynamic properties in the presence of Lorentz violation. More recently, Belchior \textit{et al.} obtained BH configurations in EKR bumblebee gravity sourced by a global monopole and analyzed their geometric and thermodynamic characteristics \cite{ref3}. Some recent investigations of optical properties, weak gravitational lensing, quasiperiodic oscillation, scalar perturbations and thermodynamics of BH solutions in EKR gravity were reported in \cite{fa1,fa2,fa3,fa4,fa5,fa6,fa7,fa8,fa9}

Besides topological defects such as global monopoles, another physically well--motivated source in general relativity is a cloud of strings, originally introduced by Letelier \cite{PSL}. A string cloud describes a spherically symmetric distribution of one--dimensional objects whose stress--energy tensor decays as $T^{t}{}_{t}=T^{r}{}_{r}\propto 1/r^{2}$, mimicking certain global defects and effectively modifying the solid angle or the mass distribution around the BH. Clouds of strings have been used as simple macroscopic models of underlying string--like structures in high--energy or early--universe contexts, and have been incorporated into BH thermodynamics and extended phase space analyses in various scenarios \cite{Ghaffarnejad2018,Toledo2018,Singh2020,Nascimento2022}. It is thus natural to ask how such matter distributions affect BH thermodynamics in the presence of Lorentz--violating sectors such as those induced by a KR field.

In this work, we investigate the thermodynamics and $P$--$V$ criticality of electrically charged AdS BHs in Einstein--Kalb--Ramond bumblebee gravity coupled to a cloud of strings. Starting from the static, spherically symmetric solutions, we derive the modified metric function in terms of the LV parameter $\ell$, the string cloud parameter $\alpha$, the electric charge $Q$, and the effective cosmological constant. Interpreting the ADM mass as enthalpy in the extended phase space, we compute the thermodynamic volume, Hawking temperature, Gibbs free energy, internal energy, and specific heat capacity, and analyze how they are deformed by the combined effects of Lorentz violation and the string cloud. We then derive the equation of state $P(T_H,v)$, identify the critical point, and show that the universal ratio $P_c v_c / T_c = 3/8$ remains unchanged, while the critical quantities $(v_c,T_c,P_c)$ acquire a nontrivial dependence on $\ell$ and $\alpha$. In addition, we study the Joule--Thomson expansion and obtain the inversion temperature, demonstrating that the ratio between the minimum inversion temperature and the critical temperature satisfies $T_i^{\min}/T_c = 1/2$, in agreement with previous results for charged AdS BHs \cite{Okcu2017,Okcu2017,Okcu2018,Mo2018PRD,Lan2018PRD,MoLi2020}. Finally, we employ the thermodynamic topology method based on a generalized free energy to characterize the topological class of the BH solutions and the nature of the phase transitions.

Our analysis thus generalizes the thermodynamic study of charged EKR bumblebee AdS BHs without additional matter \cite{ref2} and of EKR BHs with global monopoles \cite{ref3} by incorporating a cloud of strings as the source. The model provides a concrete setup in which the interplay between Lorentz--violating parameters and string--like matter content can be probed through BH thermodynamics, critical phenomena, and JT expansion, while recovering known results in the appropriate limits, such as $\alpha\to 0$ or vanishing Lorentz violation.

\section{Electrically charged AdS BHs in LV gravity with a background KR-field}
\label{sec:EKR_background}

In Ref.~\cite{ref1}, Yang \emph{et al.} constructed static, spherically symmetric neutral black hole (BH) solutions, both with and without a cosmological constant, in a gravitational theory with Lorentz violation (LV) induced by a background Kalb--Ramond (KR) field. In Ref.~\cite{ref2}, these solutions were extended by incorporating an electric charge, yielding electrically charged AdS black holes and allowing for a detailed study of their thermodynamic properties. The corresponding static and spherically symmetric electrically charged AdS BH space-time is described by the line element
\begin{align}
    ds^2 &= -h(r)\,dt^2 + \frac{dr^2}{h(r)} + r^2\left(d\theta^2 + \sin^2\theta\,d\phi^2\right), \nonumber\\
    h(r) &= \frac{1}{1-\ell} - \frac{2M}{r} + \frac{Q^2}{(1-\ell)^2\,r^2} - \frac{\Lambda_\text{eff}}{3}\,r^2,
    \label{aa1}
\end{align}
where $\Lambda_\text{eff} = \Lambda/(1-\ell)$ is the effective cosmological constant and $Q_{\rm eff} = \left|Q/(1-\ell)\right|$ is the effective electric charge. The dimensionless parameter
$\ell = \xi\,b^2/2$ encodes the LV sector, where $\xi$ is the coupling constant and $b$ is the vacuum expectation value (VEV) of the KR field; a nonzero VEV spontaneously breaks local Lorentz symmetry and deforms the metric coefficients relative to general relativity.

In Ref.~\cite{ref3}, Belchior \emph{et al.} investigated BH solutions in Einstein--Kalb--Ramond (EKR) bumblebee gravity sourced by a global monopole, again in the presence and absence of a cosmological constant. In that case, the static and spherically symmetric metric is given by
\begin{align}
    ds^2 &= -h(r)\,dt^2 + \frac{dr^2}{h(r)} + r^2\left(d\theta^2+\sin^2\theta\,d\phi^2\right), \nonumber\\
    h(r) &= \frac{1-k\,\eta^2}{1-\ell} - \frac{2M}{r} - \frac{\Lambda_\text{eff}}{3}\,r^2,
    \label{aa2}
\end{align}
where $\eta$ is the global monopole charge and $k$ parametrizes its coupling to the LV sector.

Motivated by these results, in this work we consider static and spherically symmetric AdS BH solutions in EKR bumblebee gravity where the matter source is \emph{not} a global monopole but rather a spherically symmetric cloud of strings. This allows us to probe how a string-like matter distribution, characterized by an anisotropic stress-energy tensor, modifies the LV BH geometry and its thermodynamic behavior. In addition, we generalize the neutral configurations to the electrically charged case, in close analogy with the construction of Ref.~\cite{ref2}.

A spherically symmetric cloud of strings is described by the stress-energy tensor~\cite{PSL}
\begin{equation}
    T^{\mu\nu} = \rho\,\frac{\Sigma^{\mu\beta}\,\Sigma^{\nu}{}_{\beta}}{\sqrt{-\gamma}},
    \label{aa3}
\end{equation}
where $\Sigma^{\mu\nu}$ is a bivector representing the world sheet of the strings and is given by
\begin{equation}
    \Sigma^{\mu\nu} = \epsilon^{ab}\,\frac{\partial x^{\mu}}{\partial\lambda^{a}}\,\frac{\partial x^{\nu}}{\partial\lambda^{b}},
    \label{aa4}
\end{equation}
with $\epsilon^{ab}$ the two-dimensional Levi--Civita symbol satisfying $\epsilon^{01}=-\epsilon^{10}=1$. The nonvanishing components of the stress-energy tensor for a spherically symmetric string cloud are~\cite{PSL}
\begin{align}
    T^{t}{}_{t} = T^{r}{}_{r} = \frac{\alpha}{r^2},\qquad
    T^{\theta}{}_{\theta} = T^{\phi}{}_{\phi} = 0,
    \label{aa5}
\end{align}
where $\alpha$ is a constant related to the string cloud density and controls the solid-angle deficit. Clouds of strings have been used as simple macroscopic models of underlying string-like structures in high-energy or early-universe contexts, and have been analyzed in general relativity and in several modified gravity scenarios, including Lovelock gravity, $f(R)$ gravity and nonlinear electrodynamics backgrounds~\cite{Ghaffarnejad2018_PLB785_105,Toledo2018_EPJC78_534,Toledo2019_EPJC79_117,Graca2018_CPC42_063105,Nascimento2024_Universe10_430}, and it is natural to ask how such matter distributions affect BH thermodynamics in the presence of Lorentz-violating sectors such as those induced by a KR field.

Following the methodology and field equations used in Refs.~\cite{ref1,ref2,ref3}, and replacing the global-monopole source by the string cloud described above, one arrives at the following static and spherically symmetric AdS BH solution in EKR bumblebee gravity:
\begin{equation}
    ds^2 = -f(r)\,dt^2 + \frac{dr^2}{f(r)} + r^2\left(d\theta^2+\sin^2\theta\,d\phi^2\right),
    \label{aa6}
\end{equation}
where the metric function $f(r)$ takes the form
\begin{equation}
f(r) = 
\begin{cases}
\displaystyle \frac{1 - \alpha}{1 - \ell} - \frac{2M}{r} - \dfrac{\Lambda_\text{eff}}{3}\, r^2, & \text{for } Q = 0, \\[0.7em]
\displaystyle \frac{1 - \alpha}{1 - \ell} - \frac{2M}{r} + \dfrac{Q_\text{eff}^2}{r^2} - \dfrac{\Lambda_\text{eff}}{3}\, r^2, & \text{for } Q \neq 0,
\end{cases}
\label{aa7}
\end{equation}
with $Q_\text{eff} = Q/(1-\ell)$ and $\Lambda_\text{eff} = \Lambda/(1-\ell)$ as before. The parameters $\ell$ and $\alpha$ therefore enter the geometry as effective rescalings of the constant term and the mass/charge contributions in $f(r)$: in the limit $\alpha\to0$ we recover the charged EKR AdS BH of Ref.~\cite{ref2}, whereas for $\ell\to0$ we obtain the usual cloud-of-strings black hole in general relativity.

In the absence of a cosmological constant, $\Lambda=0$, the horizon condition $f(r)=0$ leads to the radii of the inner and outer horizons of the charged BH:
\begin{align}
    r_{\pm} &= \frac{1-\ell}{1-\alpha}\,\left[1 \pm \sqrt{1-\frac{Q^2}{M^2}\,\frac{1-\alpha}{(1-\ell)^3}}\right] M, \nonumber\\
    r_{+}+r_{-} &= 2M\left(\frac{1-\ell}{1-\alpha}\right), \nonumber\\
    r_{+}-r_{-} &= 2M\left(\frac{1-\ell}{1-\alpha}\right) 
    \sqrt{1-\frac{Q^2}{M^2}\,\frac{1-\alpha}{(1-\ell)^3}}, \nonumber\\
    r_{+} r_{-} &= \frac{1-\alpha}{1-\ell}\,Q^2.
    \label{aa8}
\end{align}
The reality condition for $r_\pm$ implies an extremality bound on the charge-to-mass ratio,
\begin{equation}
\frac{Q^2}{M^2} \le (1-\ell)^3\,(1-\alpha)^{-1},
\end{equation}
showing explicitly how both the LV parameter $\ell$ and the string cloud parameter $\alpha$ modify the allowed range of $(Q,M)$ compared with the Reissner-Nordström case.

\begin{figure}[ht!]
\centering
\includegraphics[width=0.85\linewidth]{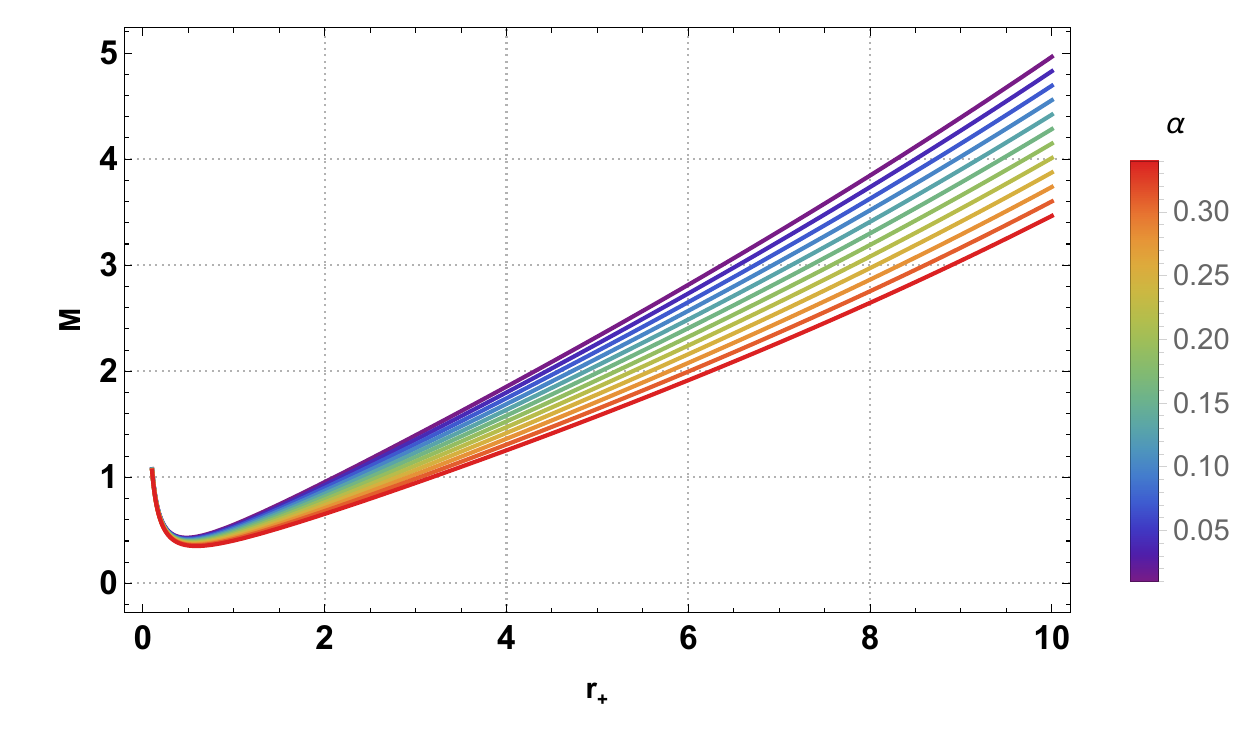}\\
(i) $\ell=-0.1$\\[0.2cm] 
\includegraphics[width=0.85\linewidth]{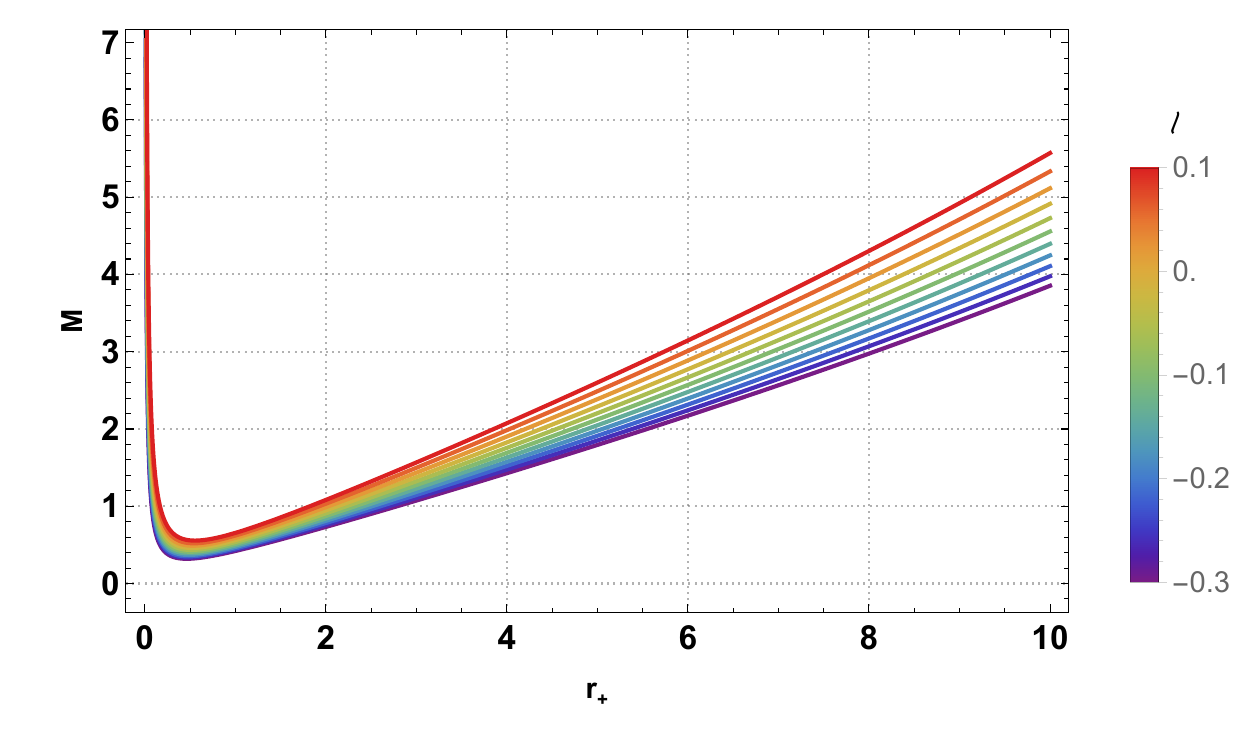}\\
(ii) $\alpha=0.1$
\caption{\footnotesize ADM mass $M$ as a function of the horizon radius $r_+$ for a charged AdS black hole in EKR bumblebee gravity with a cloud of strings. Panel~(i): fixed $\ell=-0.1$ and several values of the string cloud parameter $\alpha$. Panel~(ii): fixed $\alpha=0.1$ and several values of the LV parameter $\ell$. In both panels we set $Q=0.5$ and $\Lambda=-0.003$.}
\label{fig:ADM-mass}
\end{figure}

The dependence of the ADM mass on the horizon radius $r_+$ is illustrated in Fig.~\ref{fig:ADM-mass}. Panel~(i) shows $M(r_+)$ for fixed $\ell=-0.1$ and several values of the string cloud parameter $\alpha$, while panel~(ii) displays the effect of varying $\ell$ at fixed $\alpha=0.1$. In both cases, the mass grows monotonically with $r_+$ for sufficiently large horizons, but the small-$r_+$ behavior is sensitively deformed: increasing $\alpha$ or $\ell$ shifts the curves upward and smooths the minimum, indicating that the combined effect of the string cloud and Lorentz violation is to effectively ``dress'' the BH with an additional contribution to the energy for a given horizon size. This qualitative behavior will play an important role in the thermodynamic analysis of Sec.~\ref{sec:thermo}.

\section{Thermodynamics of BH}
\label{sec:thermo}

In this section we present the thermodynamic properties of the black hole (BH) solution introduced in Sec.~\ref{sec:EKR_background}. We derive the Hawking temperature, thermodynamic volume, Gibbs free energy, internal energy, specific heat capacity, and a modified Smarr relation, and analyse how these quantities are affected by the Kalb--Ramond (KR) LV parameter $\ell$ and the string cloud parameter $\alpha$. Throughout, we follow the extended phase space approach, in which the ADM mass $M$ is interpreted as enthalpy, the horizon area encodes the entropy, and the cosmological constant is promoted to a thermodynamic pressure~\cite{Kastor2009,Dolan2011,Cvetic2011,PhysRevD.102.044028}.

Let us first express the BH mass $M$ as a function of the horizon radius $r_+$, defined as the largest root of $f(r_+)=0$ lying inside any cosmological horizon. Using the charged metric function in Eq.~\eqref{aa7}, one finds
\begin{equation}
    M = \frac{r_{+}}{2}\left[\frac{1 - \alpha}{1 - \ell}
    + \frac{Q^2}{(1-\ell)^2\,r^2_{+}}
    - \frac{\Lambda}{3\,(1-\ell)} \, r^2_{+}\right].
    \label{bb1}
\end{equation}
In the extended phase space, the cosmological constant is related to the thermodynamic pressure $P$ as~\cite{ref2,PhysRevD.102.044028}
\begin{equation}
\Lambda = -8\,\pi\,(1-\ell)\,P.
\label{bb2}
\end{equation}
Substituting Eq.~\eqref{bb2} into Eq.~\eqref{bb1}, the mass (enthalpy) can be written in the standard $P$–$V$ form
\begin{equation}
    M = \frac{r_{+}}{2}\left[\frac{1 - \alpha}{1 - \ell}
    + \frac{Q^2}{(1-\ell)^2\,r^2_{+}}
    + \frac{8\,\pi}{3} \, r^2_{+}\,P\right].
    \label{bb3}
\end{equation}
In the limit $\alpha \to 0$, the contribution of the string cloud disappears, and Eq.~\eqref{bb3} reduces to the enthalpy of the charged EKR AdS black hole of Ref.~\cite{ref2}.

\begin{center}
    {\bf A.\,Thermodynamic volume}
\end{center}

In the extended phase space of black hole thermodynamics, with $\Lambda$ treated as a thermodynamic pressure $P$, the thermodynamic volume $V$ is defined as the quantity conjugate to $P$ in the first law,
\[
V = \left(\frac{\partial M}{\partial P}\right)_{S,Q,\alpha,\ell}.
\]
Using Eq.~\eqref{bb3} we obtain
\begin{equation}
    V = \left(\frac{\partial M}{\partial P}\right)_{\alpha,\ell,Q}
    = \frac{4\,\pi}{3}\, r^3_{+},
    \label{bb4}
\end{equation}
which coincides with the geometric volume inside the horizon and is identical to the thermodynamic volume of the Schwarzschild and Reissner--Nordström AdS black holes~\cite{Cvetic2011}. The LV parameter $\ell$ and the string parameter $\alpha$ do not modify $V$ explicitly, but they do affect $r_+$ for fixed $(M,Q,P)$.

\begin{center}
{\bf B.\,Hawking temperature}
\end{center}

The Hawking temperature is a fundamental thermodynamic quantity associated with black hole radiation, first derived by Hawking in the context of quantum field theory in curved spacetime~\cite{Hawking1975}. It is given by
\begin{equation}
T_H = \frac{\kappa}{2\pi},
\label{Haw}    
\end{equation}
where 
\begin{equation}
\kappa = -\frac{1}{2}\,\frac{\partial_r g_{tt}}{\sqrt{-g_{tt}\,g_{rr}}}\Big{|}_{r=r_{+}}
\label{Haw2}    
\end{equation}
is the surface gravity at the black hole horizon. This result shows that black holes are not completely black, but emit thermal radiation, establishing a deep connection between gravitation, thermodynamics, and quantum theory.

In our model, the Hawking temperature is obtained as 
\begin{align}
    T_H=\frac{1}{4\pi\,r_{+}}\left[\frac{1 - \alpha}{1 - \ell}
    - \frac{Q^2}{(1 - \ell)^2\, r_+^2}
    + 8\pi P\,r^2_+\right].
    \label{bb5}
\end{align}

From Eq.~\eqref{bb5} it is clear that $T_H$ is controlled by four independent parameters: the LV parameter $\ell$, the string cloud parameter $\alpha$, the charge $Q$, and the pressure $P$. For fixed $(Q,P)$, increasing $\alpha$ effectively weakens the constant term in $f(r)$ and tends to reduce the temperature at a given $r_+$, while variations in $\ell$ rescale both the constant and charge terms, leading to nontrivial shifts in the extremal radius and in the location of the temperature minimum.

\begin{figure}[ht!]
\centering
\includegraphics[width=0.85\linewidth]{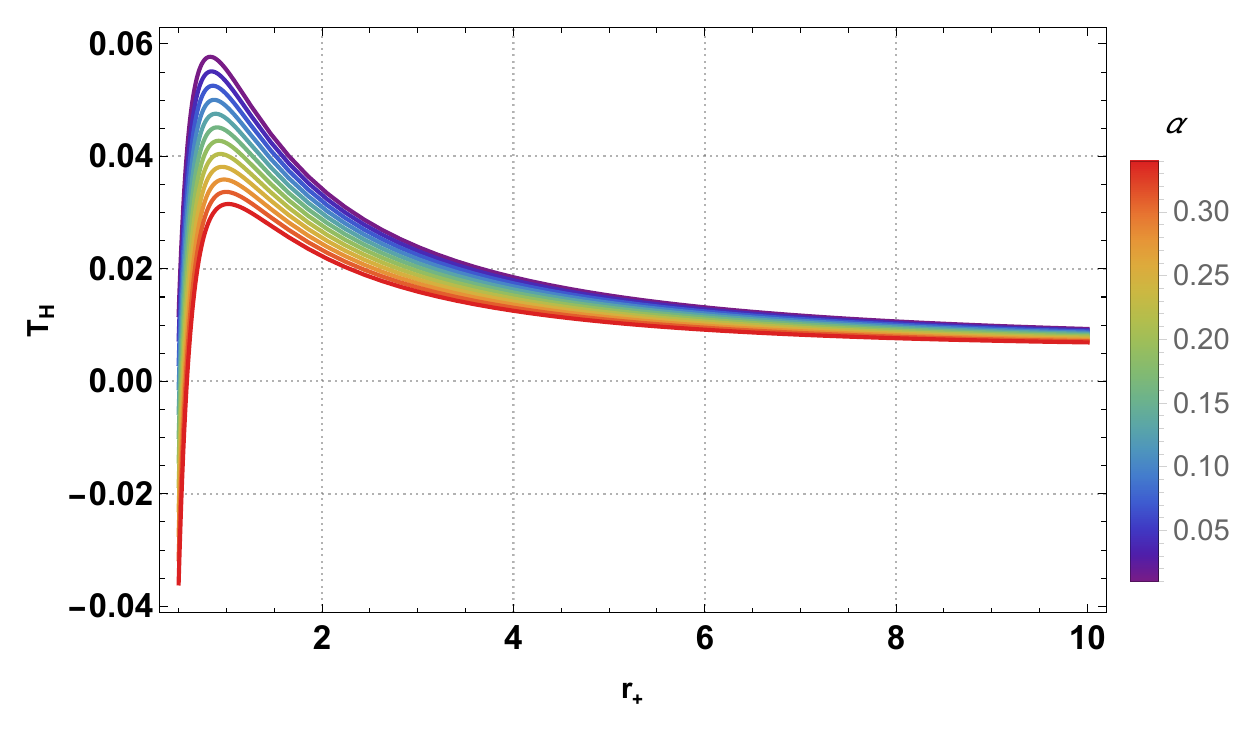}\\
(i) $\ell=-0.1$\\
\includegraphics[width=0.85\linewidth]{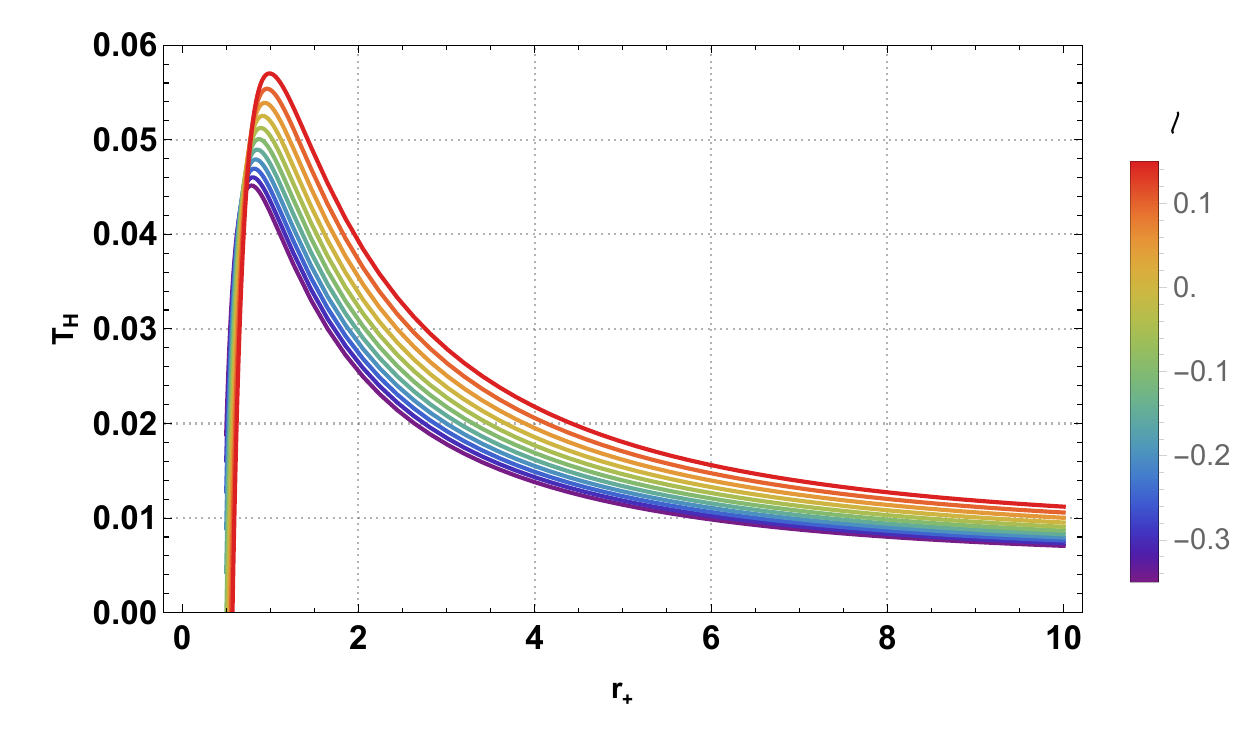}\\
(ii) $\alpha=0.1$
\caption{\footnotesize
Hawking temperature $T_H$ as a function of the horizon radius $r_+$ for a charged AdS black hole in EKR bumblebee gravity with a cloud of strings. Panel~(i): fixed $\ell=-0.1$ and different values of the string cloud parameter $\alpha$. Panel~(ii): fixed $\alpha=0.1$ and different values of the LV parameter $\ell$. In both cases we set $Q=0.5$ and $\Lambda=-0.003$. The deformation of the $T_H(r_+)$ curves relative to the Reissner--Nordström--AdS case illustrates the impact of $\ell$ and $\alpha$ on the small- and large-black-hole branches.}
\label{fig:Hawking}
\end{figure}

Figure~\ref{fig:Hawking} displays the Hawking temperature as a function of the horizon radius $r_+$ for representative values of $(\ell,\alpha)$, with $Q=0.5$ and $\Lambda=-0.003$. In each panel one observes the characteristic AdS behavior: $T_H$ vanishes at an extremal radius, increases to a maximum (or exhibits a monotonic rise beyond a minimum, depending on parameters), and grows linearly for large $r_+$ due to the $8\pi P r_+^2$ term. The position and height of the minimum are sensitive to both $\ell$ and $\alpha$, indicating that Lorentz violation and the string cloud jointly modulate the crossover between small and large black hole branches. This structure will be reflected in the sign changes and divergences of the specific heat in Sec.~\ref{sec:thermo_heat}.

\begin{center}
    {\bf C.\,Gibbs free energy and internal energy}
\end{center}

In the extended phase space, the Gibbs free energy $F$ of a black hole is defined as
\begin{equation}\label{bb6}
F = M - T_H\,S,
\end{equation}
where $M$ is the black hole mass (interpreted as enthalpy), $T_H$ is the Hawking temperature, and $S$ is the entropy associated with the BH horizon. The dependence of $F$ on the temperature or on the horizon radius encodes information about global thermodynamic stability and possible first-order phase transitions~\cite{Chamblin1999,KubiznakMann2012}.

The entropy of the black hole can be obtained as follows:
\begin{equation}
S = \frac{A}{4} = \pi r_+^2,\qquad 
A(r_+) = \int\!\!\int \sqrt{g_{\theta\theta}\,g_{\phi\phi}}\, d\theta\, d\phi
= 4 \pi r^2_{+},
\label{bb7a}
\end{equation}
which is exactly the Bekenstein--Hawking entropy formula, where $r_+$ is the radius of the event horizon. Thus, the LV parameter and the string cloud do not modify the area law: their effects enter solely through the relation between $r_+$ and the thermodynamic variables.

Using Eqs.~\eqref{bb3} and~\eqref{bb5}, we find the Gibbs free energy
\begin{equation}
F = \frac{r_+}{4}\left[
\frac{1 - \alpha}{1 - \ell}+ \frac{3\,Q^2}{(1 - \ell)^2\,r^2_+}- \frac{8}{3}\pi r_+^2 P\right].
\label{bb8}
\end{equation}

\begin{figure}[ht!]
    \centering
    \includegraphics[width=0.85\linewidth]{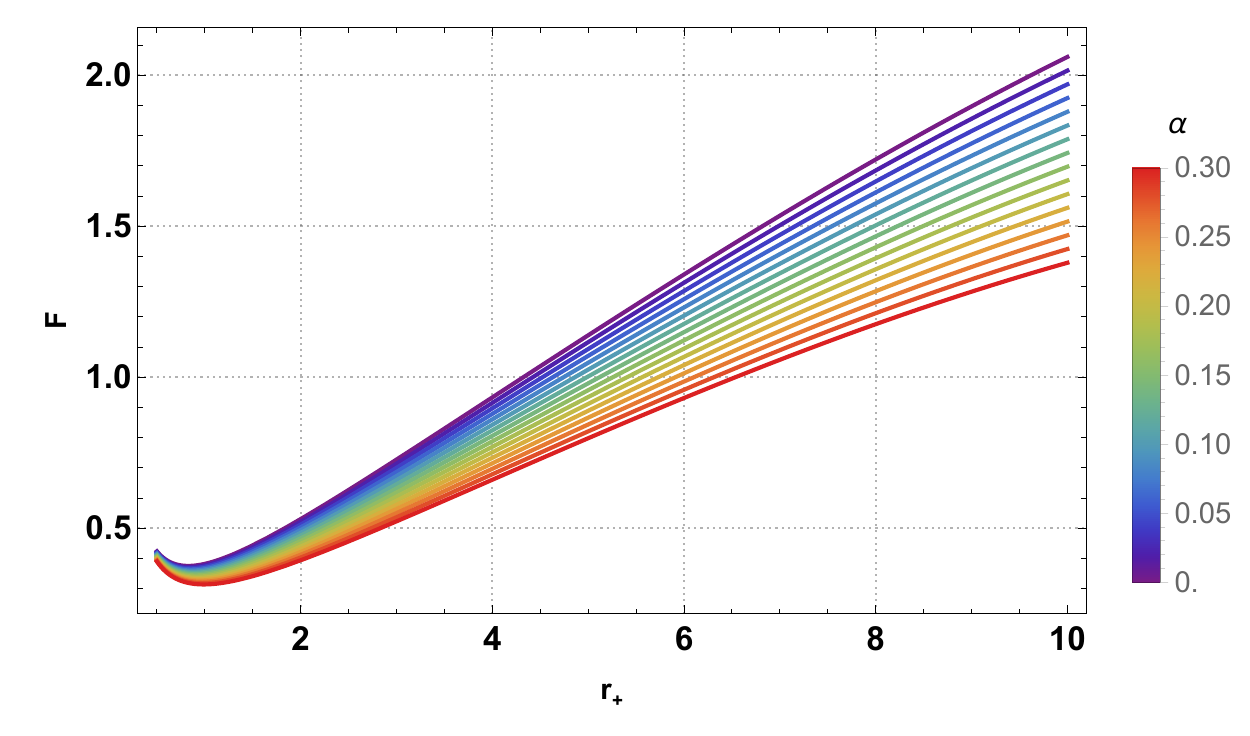}\\
    (i) $\ell=-0.1$\\
    \includegraphics[width=0.85\linewidth]{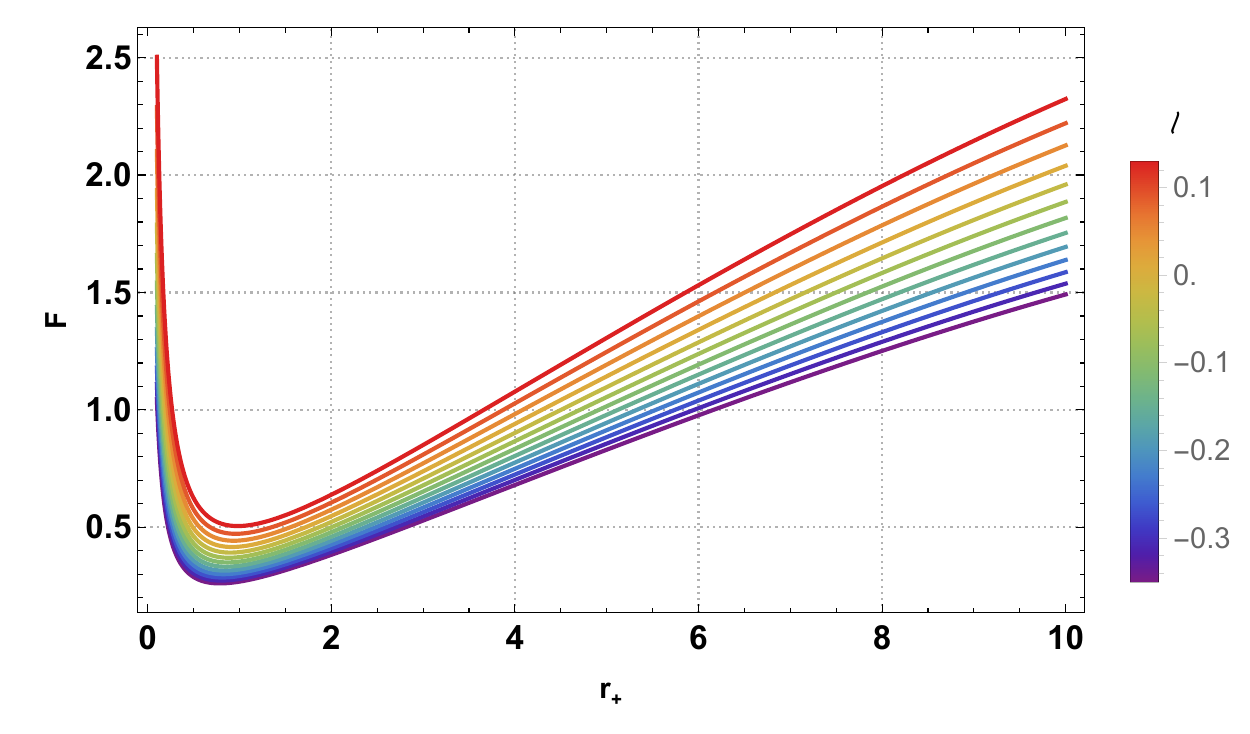}\\
    (ii) $\alpha=0.1$
    \caption{\footnotesize
    Gibbs free energy $F$ as a function of the horizon radius $r_{+}$ for a charged AdS black hole in EKR bumblebee gravity with a cloud of strings. Panel~(i): fixed $\ell=-0.1$ and different values of $\alpha$. Panel~(ii): fixed $\alpha=0.1$ and different values of $\ell$. Here $Q=0.5$ and $\Lambda=-0.003$. The shifts in $F(r_+)$ show how the LV and string parameters influence the relative thermodynamic preference of different horizon sizes.}
    \label{fig:Gibbs-energy}
\end{figure}

The behavior of $F$ as a function of $r_+$ is illustrated in Fig.~\ref{fig:Gibbs-energy}. For fixed $(Q,P)$ and $\ell=-0.1$ [panel~(i)], increasing the string parameter $\alpha$ shifts the $F(r_+)$ curves upward and can delay the crossing of $F=0$, indicating that the presence of the string cloud tends to make small black holes less thermodynamically favored. Conversely, for fixed $\alpha=0.1$ [panel~(ii)], varying $\ell$ deforms the balance between the constant, charge, and $P r_+^2$ terms, effectively mimicking a rescaling of the cosmological constant and charge contributions. Although the figure is plotted as $F$ versus $r_+$ rather than $F$ versus $T_H$, the qualitative shape still reflects the underlying small/large-black-hole structure that, at fixed $P$, is associated with the Van der Waals-like phase behavior discussed in Sec.~\ref{sec:PV}.

\begin{figure}[ht!]
\centering
\includegraphics[width=0.85\linewidth]{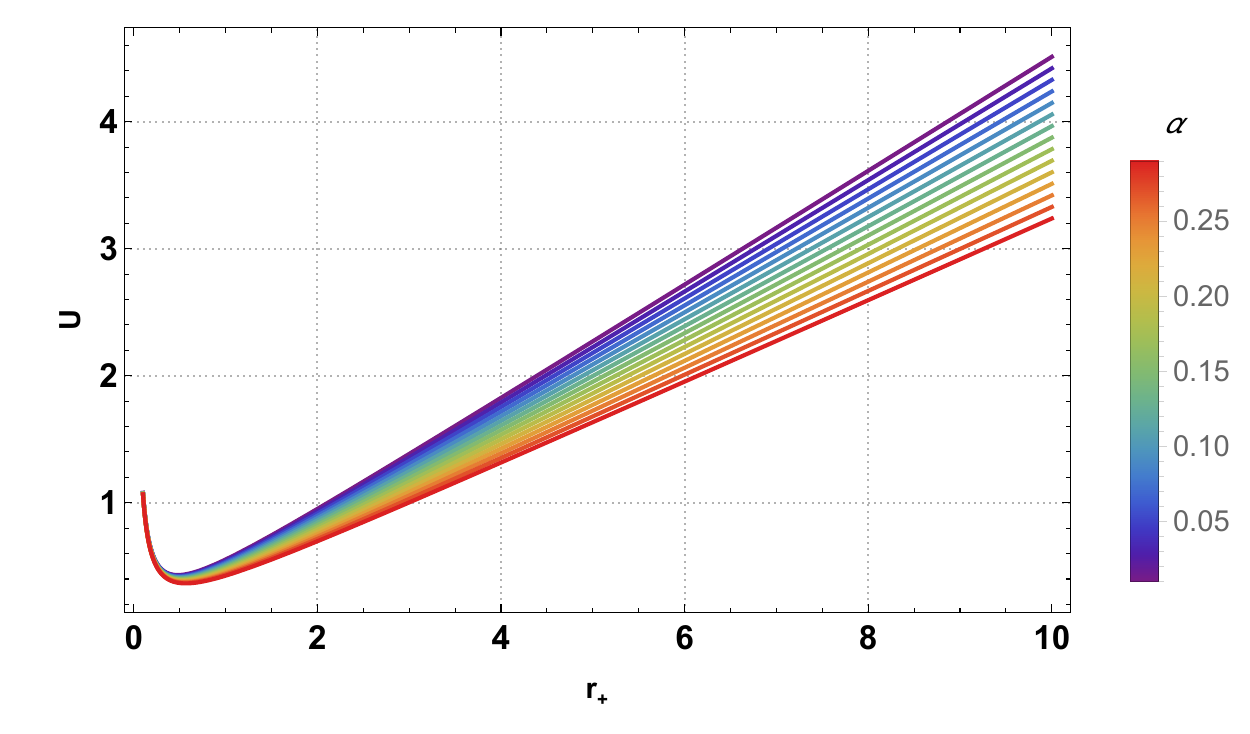}\\
(i) $\ell=-0.1$\\
\includegraphics[width=0.85\linewidth]{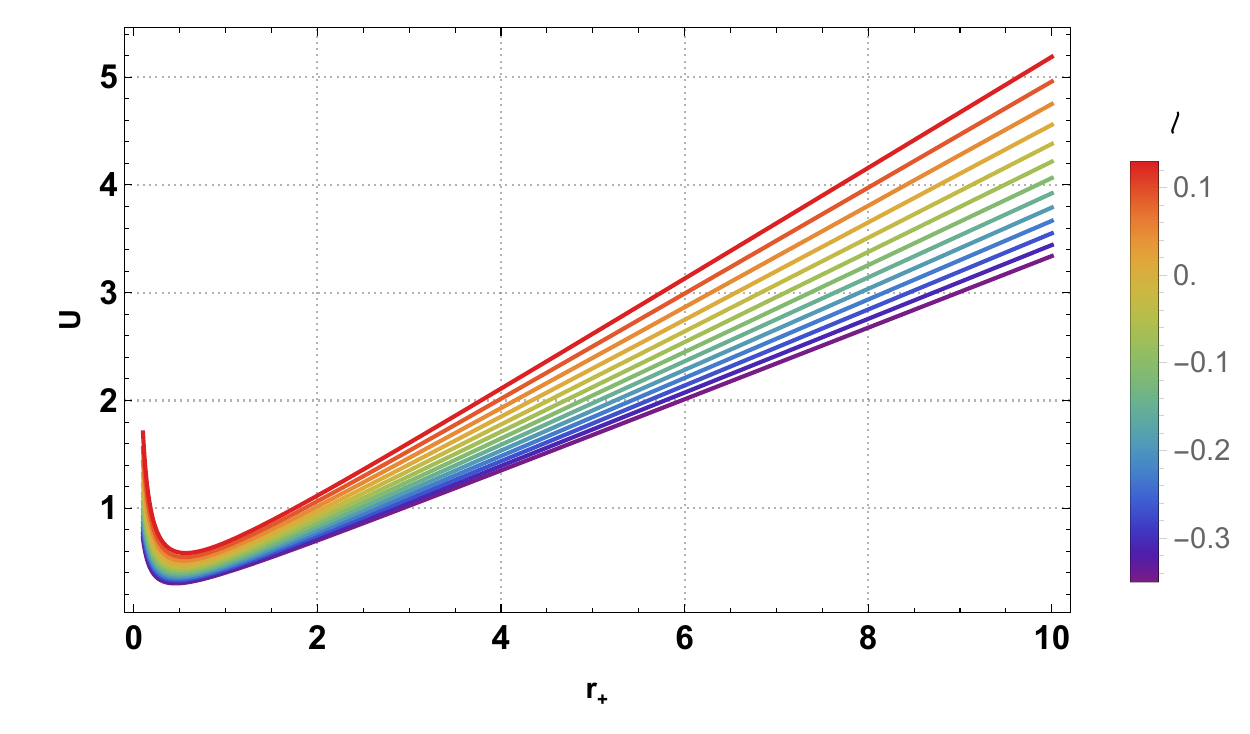}\\
(ii) $\alpha=0.1$
\caption{\footnotesize
Internal energy $U$ as a function of the horizon radius $r_{+}$ for $Q=0.5$. Panel~(i): fixed $\ell=-0.1$ and varying $\alpha$. Panel~(ii): fixed $\alpha=0.1$ and varying $\ell$. The presence of the string cloud and Lorentz violation modifies both the slope and magnitude of $U(r_+)$, effectively ``dressing'' the black hole energy for a given horizon size.}
\label{fig:intern-energy}
\end{figure}

The internal energy $U$ of the BH is obtained from the enthalpy by subtracting the $P V$ term,
\begin{equation}
    U = M - P\,V
    = \frac{r_+}{2}\left[
    \frac{1 - \alpha}{1 - \ell}
    + \frac{Q^2}{(1 - \ell)^2\, r^2_+}\right].
    \label{bb9a}
\end{equation}
In the limit $\alpha \to 0$, Eq.~\eqref{bb9a} reproduces the internal energy of the charged EKR AdS BH obtained in Ref.~\cite{ref2}. As a function of $r_+$, $U$ is monotonic for the parameter ranges considered, but its slope and magnitude are controlled by $\ell$ and $\alpha$, as displayed in Fig.~\ref{fig:intern-energy}. The string cloud increases the effective energy stored near the horizon, while the LV parameter modifies the relative weight of the constant and Coulombic contributions.

\begin{center}
{\bf D.\,Specific heat capacity}
\label{sec:thermo_heat}
\end{center}

The specific heat capacity of a black hole at constant pressure (or other fixed parameters) is a key quantity that characterizes its thermodynamic stability. It is defined as
\begin{equation}
C_{\rm heat} = T_H\, \left( \frac{\partial S}{\partial T} \right)_{\!P, Q, \alpha, \ell},
\label{bb9}
\end{equation}
where $T_H$ is the Hawking temperature and $S$ is the entropy.

A positive specific heat indicates local thermodynamic stability, since the system can absorb heat without undergoing runaway changes in temperature, while a negative specific heat signals instability and is typically associated with small-black-hole branches or evaporation processes~\cite{Davies1977,BrownMann1994}. In extended phase space thermodynamics, where $P$ and $V$ are treated as thermodynamic variables, the specific heat can exhibit discontinuities or divergences that signal phase transitions analogous to those in Van der Waals fluids~\cite{Chamblin1999,KubiznakMann2012}.

The above definition can be rewritten as 
\begin{equation}\label{bb10}
C_{\rm heat} = T_H\left( \frac{\partial S}{\partial T} \right)
= T_H\left( \frac{\partial S}{\partial r_{+}} \right)\Bigg/\left( \frac{\partial T_H}{\partial r_{+}}\right),
\end{equation}
which is convenient because both $S$ and $T_H$ are explicit functions of $r_+$.

Using the Hawking temperature $T_H$ given in Eq.~\eqref{bb5}, we find the specific heat capacity as
\begin{equation}
C_{\rm heat} = 2\pi\,r^2_{+}\,
\frac{\displaystyle \frac{1 - \alpha}{1 - \ell}
- \frac{Q^2}{(1 - \ell)^2\, r_+^2}
+ 8 \pi P r^2_+}
{\displaystyle -\frac{1 - \alpha}{1 - \ell}+ \frac{3\,Q^2}{(1 - \ell)^2\, r_+^2}+ 8 \pi P r^2_{+}}.
\label{bb11}
\end{equation}
The sign and divergences of $C_{\rm heat}$ are controlled by the denominator of Eq.~\eqref{bb11}. For parameter choices that admit two distinct positive roots of $\partial T_H/\partial r_+ = 0$, one finds three regions in $r_+$: a small-black-hole branch with negative specific heat, an intermediate branch where $C_{\rm heat}$ diverges and changes sign, and a large-black-hole branch with positive specific heat. This pattern is characteristic of Van der Waals-like behavior and underlies the $P$–$V$ criticality discussed in Sec.~\ref{sec:PV}~\cite{Chamblin1999,KubiznakMann2012}.

\begin{figure}[ht!]
\centering
\includegraphics[width=0.85\linewidth]{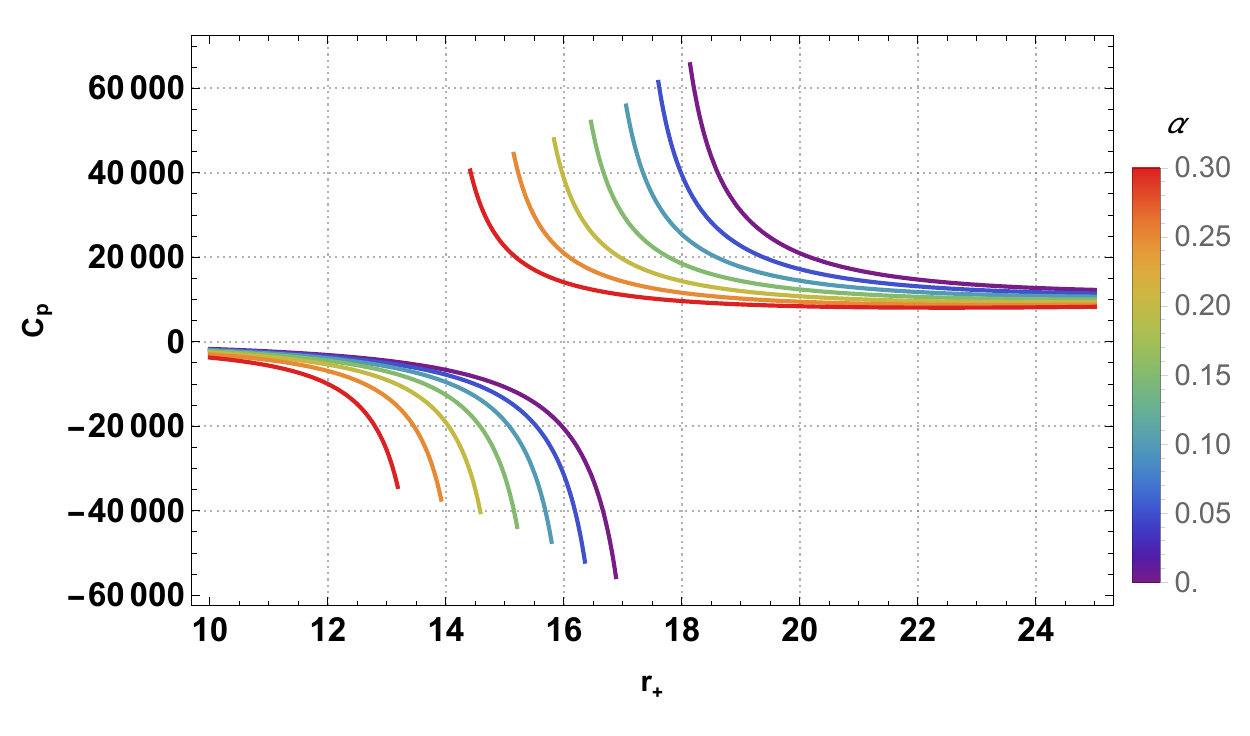}\\
(i) $\ell=-0.1$ \\
\includegraphics[width=0.85\linewidth]{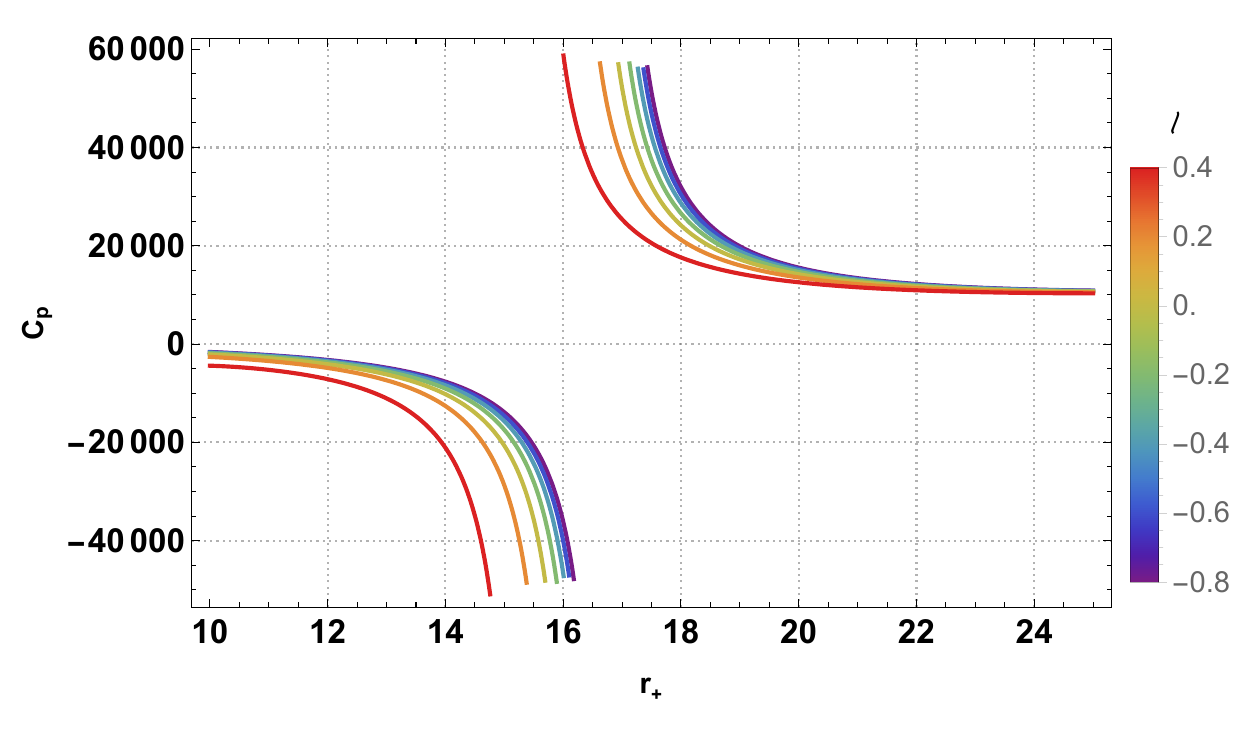}\\
(ii) $\alpha=0.1$
\caption{\footnotesize
Specific heat capacity $C_{\rm heat}$ at constant pressure as a function of the horizon radius $r_{+}$, for $\Lambda=-0.003$ and $Q=3$. Panel~(i): fixed $\ell=-0.1$ and varying $\alpha$. Panel~(ii): fixed $\alpha=0.1$ and varying $\ell$. The zeros and divergences of $C_{\rm heat}$ delimit stable ($C_{\rm heat}>0$) and unstable ($C_{\rm heat}<0$) black hole branches, and their dependence on $\ell$ and $\alpha$ encodes the influence of Lorentz violation and the string cloud on local thermodynamic stability.}
\label{fig:heat-capacity}
\end{figure}

Figure~\ref{fig:heat-capacity} presents $C_{\rm heat}$ as a function of $r_+$ for $\Lambda=-0.003$ and $Q=3$. For fixed $\ell=-0.1$ [panel~(i)], increasing $\alpha$ modifies both the location and height of the divergences, shrinking or enlarging the stable large-$r_+$ region depending on the string density. For fixed $\alpha=0.1$ [panel~(ii)], varying $\ell$ similarly shifts the radii where $C_{\rm heat}$ changes sign, reflecting the role of Lorentz violation in controlling the local thermodynamic stability of the BH. Regions where $C_{\rm heat}>0$ correspond to thermodynamically stable configurations, while regions with $C_{\rm heat}<0$ are unstable.

\begin{center}
    {\bf E.\,Modified first law and Smarr relation}
\end{center}

Given the enthalpy $M(r_+)$ in Eq.~\eqref{bb3}, we can now write the first law of black hole thermodynamics including the string cloud parameter $\alpha$ as an additional thermodynamic variable. Treating $\alpha$ as an intensive parameter, the first law reads
\begin{equation}
dM = T_H\, dS + \Phi_Q\,dQ_{\rm eff} + V\, dP + \Phi_{\alpha}\, d\alpha,
\label{gg1}
\end{equation}
where $T_H$ is the Hawking temperature, $S$ the entropy, $V$ the thermodynamic volume conjugate to the pressure $P$, $\Phi_Q$ the electric potential conjugate to the effective charge $Q_{\rm eff}$, and $\Phi_{\alpha}$ the conjugate quantity associated with the string cloud parameter $\alpha$~\cite{Kastor2009,Dolan2011}. These are given by
\begin{align}
\Phi_Q &= \left(\frac{\partial M}{\partial Q_{\rm eff}}\right)_{S,P,\alpha}= \frac{Q}{(1-\ell)\,r_{+}}
= \frac{Q_{\rm eff}}{r_{+}}, \nonumber\\
\Phi_{\alpha} &= \left(\frac{\partial M}{\partial \alpha}\right)_{S,P,Q}= -\frac{r_+}{2 (1-\ell)}.
\label{gg2}
\end{align}    

Combining $T_H$, $S$, $V$, and the potentials $\Phi_Q$ and $\Phi_{\alpha}$, one obtains the following generalized Smarr relation~\cite{Kastor2009,Dolan2011}
\begin{equation}
2 (T_H S - P V) + \Phi_Q\,Q_{\rm eff} + \Phi_{\alpha}\,\alpha= M - \frac{\alpha r_{+}}{2 (1-\ell)} \equiv \mathcal{M},\label{gg3}
\end{equation}
where $\mathcal{M}$ can be interpreted as an effective mass parameter once the contribution of the string cloud is separated out. In the limit $\alpha=0$, corresponding to the absence of the string cloud, Eq.~\eqref{gg3} reduces to
\[
M = 2 (T_H S - P V) + \Phi_Q\,Q_{\rm eff},
\]
which is analogous to the Smarr formula for Reissner-Nordström-AdS black holes, now written in terms of the LV-dressed charge $Q_{\rm eff}$. This confirms the consistency of the extended thermodynamic description in the presence of both Lorentz violation and a string cloud.

\section{$P$--$V$ criticality of BH}
\label{sec:PV}

In extended black hole thermodynamics the cosmological constant is promoted to a thermodynamic pressure and the ADM mass is interpreted as enthalpy rather than internal energy (see, e.g., Refs.~\cite{KubiznakMann2012,KubiznakMannTeo2017}). In the present Lorentz-violating setup, the relation between $\Lambda$ and $P$ is modified according to Eq.~\eqref{bb2}, $\Lambda=-8\pi(1-\ell)P$, but the overall structure of the phase space remains analogous to the standard AdS case. A paradigmatic example is the Reissner--Nordström--AdS black hole, whose equation of state $P(T,v)$, written in terms of the specific volume $v=2r_+$, closely resembles that of a Van der Waals fluid~\cite{KubiznakMann2012,ZhangCaiYu2015}. Solving the criticality conditions $(\partial P/\partial v)_T=0$ and $(\partial^2P/\partial v^2)_T=0$ yields a critical point with universal fluid-like behaviour: the ratio $P_c v_c/T_c = 3/8$ matches that of the Van der Waals gas, and the critical exponents take their classical mean-field values~\cite{KubiznakMann2012,ZhangCaiYu2015}. In our model, we will see that the presence of the Kalb--Ramond LV parameter $\ell$ and the string cloud parameter $\alpha$ deforms the critical scales $(v_c,T_c,P_c)$ but preserves the same universal ratio.

Starting from the Hawking temperature \eqref{bb5}, we can rewrite it as a function of the pressure $P$,
\begin{equation}
    T_H = \frac{1}{4\pi\,r_{+}}\left[\frac{1 - \alpha}{1 - \ell}
- \frac{Q^2}{(1 - \ell)^2\, r_+^2}
+8\pi\,P\,r^2_+\right].
\label{gg4}
\end{equation}
Inverting this relation we obtain the equation of state $P(T_H,r_+)$,
\begin{align}
    P &= \frac{1}{8\pi\,r^2_{+}}\left[4\pi\,r_{+}\,T_H
    -\frac{1 - \alpha}{1 - \ell}
    +\frac{Q^2}{(1 - \ell)^2\, r_+^2}\right] \nonumber\\
    &= \frac{T_H}{2\,r_{+}}-\frac{\eta}{4\,r^2_{+}}
    +\frac{\zeta}{16\,r^4_{+}},
    \label{gg5}
\end{align}
where 
\begin{equation}
\eta=\frac{1}{2\pi}\,\frac{1 - \alpha}{1 - \ell},
\qquad 
\zeta=\frac{2\,Q^2}{\pi\,(1-\ell)^2}.
\label{constant}    
\end{equation}

Introducing the specific volume $v=2r_+$, Eq.~\eqref{gg5} becomes
\begin{equation}
P = \frac{T_H}{v} - \frac{\eta}{v^2} + \frac{\zeta}{v^4},
\label{gg6}
\end{equation}
which has the same functional form as the RN--AdS equation of state, with the effective parameters $\eta(\ell,\alpha)$ and $\zeta(\ell,Q)$ encoding the impact of Lorentz violation and the string cloud. For $\alpha=0$ and $\ell=0$, one recovers the standard RN--AdS result~\cite{KubiznakMann2012,ZhangCaiYu2015}.

The critical point $(v_c, T_c, P_c)$ is obtained by imposing
\begin{align}
    &\frac{\partial P}{\partial v}\Big|_{T_H}=0,
    \label{gg7}\\
    &\frac{\partial^2 P}{\partial v^2}\Big|_{T_H}=0,
    \label{gg8}
\end{align}
which eliminate the inflection point of the isotherms in the $P$--$v$ plane. Using Eqs.~\eqref{gg6}–\eqref{gg8}, one finds
\begin{align}
v_c &= \sqrt{\frac{6\,\zeta}{\eta}}
= 2\,\sqrt{\frac{6}{(1-\ell)(1-\alpha)}}\,Q,
\nonumber\\
T_c &= \frac{4\,\eta}{3}\,\sqrt{\frac{\eta}{6\,\zeta}}
= \frac{(1-\alpha)^{3/2}}
{3\pi\,\sqrt{6}\,Q\,(1-\ell)^{1/2}},
\nonumber\\
P_c &= \frac{\eta^2}{12\,\zeta}
= \frac{(1-\alpha)^2}{96\pi \,Q^2}.
\label{gg9}
\end{align}
The corresponding critical ratio is
\begin{equation}
    R_{\rm critical}=\frac{P_c\,v_c}{T_c}=\frac{3}{8},
    \label{gg10}
\end{equation}
which coincides exactly with the Van der Waals value, showing that the LV parameter $\ell$ and the string cloud parameter $\alpha$ do not modify the universality class of the phase transition. Interestingly, $P_c$ is independent of $\ell$ and depends only on $\alpha$ and $Q$, while both $v_c$ and $T_c$ are sensitive to the Lorentz-violating background and to the string cloud.

The critical radius $r_c=v_c/2$ determines the critical thermodynamic volume,
\begin{equation}
    V_c = \frac{4\pi}{3} r^3_c
    = 8 \sqrt{6}\,\pi\,Q^3\,(1-\alpha-\ell+\alpha \ell)^{-3/2}.
\end{equation}
For comparison, the RN--AdS critical parameters~\cite{KubiznakMann2012,ZhangCaiYu2015} are
\begin{align}
v_c = 2\,\sqrt{6}\,Q,\;
T_c = \frac{1}{3\pi\,\sqrt{6}\,Q},\; 
P_c = \frac{1}{96\pi \,Q^2},\; 
V_c = 8 \sqrt{6} \pi Q^3,
\label{gg11}
\end{align}
which are recovered from Eq.~\eqref{gg9} in the limit $\alpha\to 0$ and $\ell\to 0$.

The dependence of the critical size and temperature on $(\alpha,\ell)$ is summarized in Tables~\ref{tab:1} and~\ref{tab:2}, respectively, while Table~\ref{tab:3} collects the values of $P_cQ^2$ for several $\alpha$, highlighting the RN--AdS case in bold. As expected from Eq.~\eqref{gg9}, $v_c/Q$ increases monotonically with both $\alpha$ and $\ell$, showing that the presence of the string cloud and Lorentz violation enlarges the effective critical size of the black hole. Conversely, $T_c Q$ decreases with increasing $\alpha$ (for fixed $\ell$) and increases with $\ell$ (for fixed $\alpha$), indicating that the string cloud tends to lower the critical temperature, while a positive LV parameter pushes it upward. The critical pressure $P_c Q^2$ decreases as $\alpha$ grows and remains independent of $\ell$, in agreement with the analytical expression in Eq.~\eqref{gg9}. Similar deformations of the critical scales driven by string-cloud parameters have also been reported in other AdS black hole backgrounds, including Gauss--Bonnet massive gravity and nonsingular metrics~\cite{Ranjbari2020,Sood2022}.

\begin{table}[ht!]
\centering
\caption{Numerical values of the critical size $v_c / Q$ for different $\alpha$ and $\ell$ and a comparison with RN--AdS.}
\vspace{0.2cm}
\begin{tabular}{|c|c|c|c|c|c|c|}
\hline
$\alpha \backslash \ell$ & -0.3 & -0.2 & -0.1 & 0 & 0.1 & 0.2 \\
\hline
0     & 4.2967 & 4.4721 & 4.6710 & {\bf 4.8989} & 5.1640 & 5.4772 \\
0.05  & 4.4083 & 4.5883 & 4.7923 & 5.0263  & 5.2981 & 5.6195 \\
0.10  & 4.5291 & 4.7141 & 4.9237 & 5.1640  & 5.4433 & 5.7735 \\
0.15  & 4.6604 & 4.8507 & 5.0664 & 5.3137  & 5.6011 & 5.9409 \\
0.20  & 4.8038 & 5.0000 & 5.2223 & 5.4772  & 5.7735 & 6.1237 \\
0.25  & 4.9614 & 5.1640 & 5.3936 & 5.6569  & 5.9629 & 6.3246 \\
\hline
\end{tabular}
\label{tab:1}
\end{table}

\begin{table}[ht!]
\centering
\caption{Numerical values of the critical temperature $T_c\,Q$ for different $\alpha$ and $\ell$ and a comparison with RN--AdS.}
\vspace{0.2cm}
\begin{tabular}{|c|c|c|c|c|c|c|}
\hline
$\alpha \backslash \ell$ & -0.3 & -0.2 & -0.1 & 0 & 0.1 & 0.2 \\
\hline
0     & 0.03799 & 0.03954 & 0.04130 & {\bf 0.04305} & 0.04566 & 0.04843 \\
0.05  & 0.03518 & 0.03661 & 0.03824 & 0.04011 & 0.04228 & 0.04484 \\
0.10  & 0.03244 & 0.03376 & 0.03526 & 0.03698 & 0.03898 & 0.04135 \\
0.15  & 0.02977 & 0.03099 & 0.03237 & 0.03395 & 0.03578 & 0.03795 \\
0.20  & 0.02718 & 0.02829 & 0.02955 & 0.03099 & 0.03267 & 0.03465 \\
0.25  & 0.02468 & 0.02568 & 0.02683 & 0.02813 & 0.02966 & 0.03146 \\
\hline
\end{tabular}
\label{tab:2}
\end{table}

\begin{table}[ht!]
\centering
\caption{Numerical values of the critical pressure $P_c\,Q^2$ for different $\alpha$ and a comparison with RN--AdS.}
\vspace{0.2cm}
\begin{tabular}{|c|c|}
\hline
$\alpha$ & $P_c Q^2$ \\
\hline
0    & {\bf 0.00331} \\
0.05  & 0.00299 \\
0.10  & 0.00269 \\
0.15  & 0.00240 \\
0.20  & 0.00212 \\
0.25  & 0.00187 \\
\hline
\end{tabular}
\label{tab:3}
\end{table}

\begin{figure}[ht!]
\centering
\includegraphics[width=0.8\linewidth]{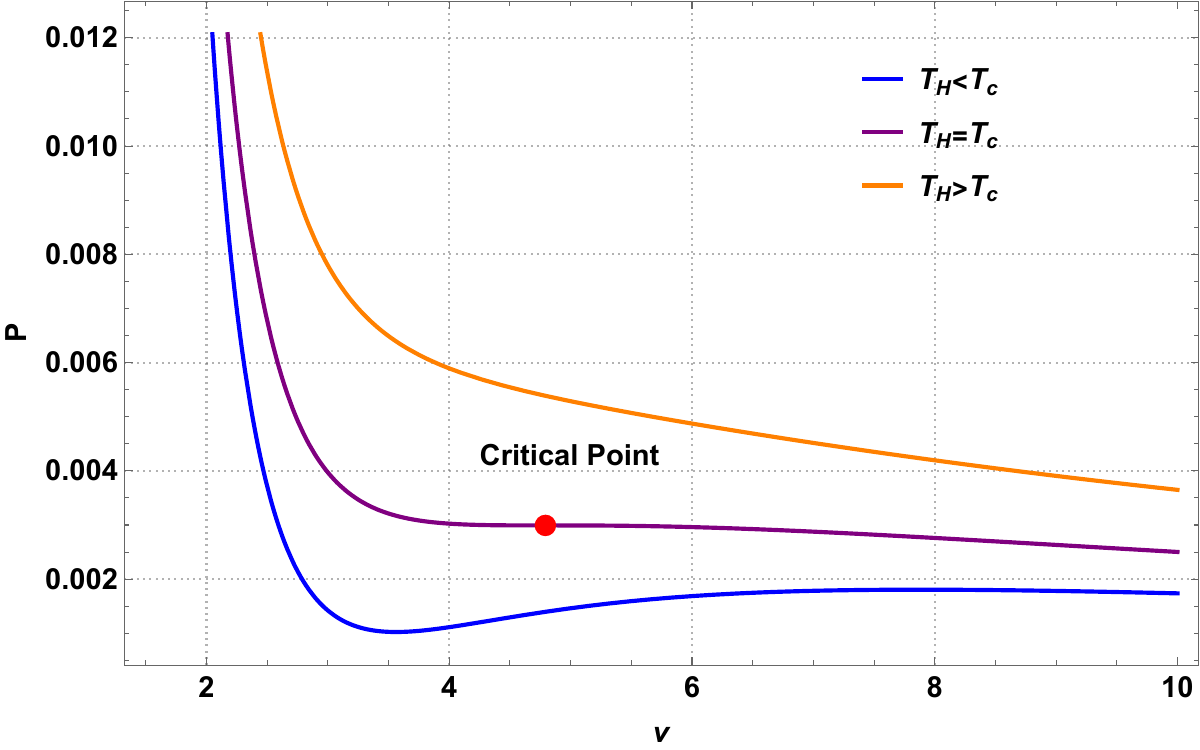}\\
(i) $\alpha=0.05$ \\
\includegraphics[width=0.8\linewidth]{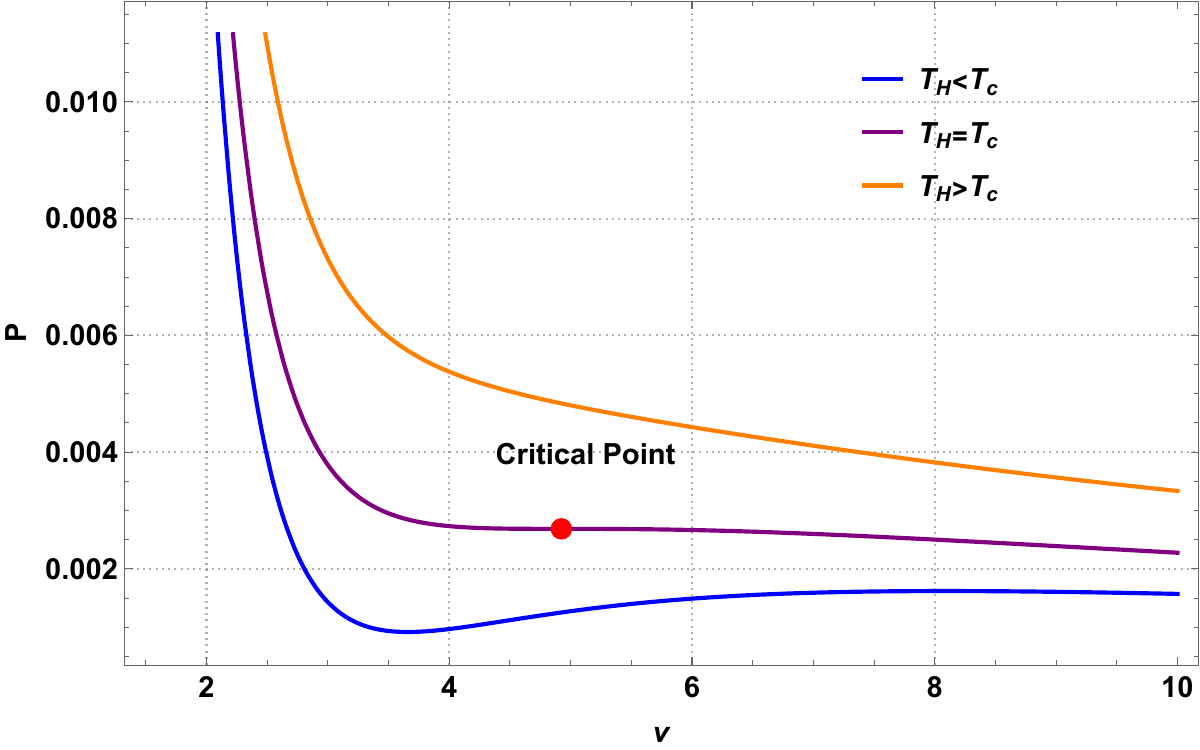}\\
(ii) $\alpha=0.1$
\caption{\footnotesize
$P$--$v$ diagrams for the charged AdS black hole in EKR bumblebee gravity with a cloud of strings, at fixed $\ell=-0.1$ and $Q=1$. Panel~(i): $\alpha=0.05$ with critical point $(v_c=4.79234,\,T_c = 0.0382422,\,P_c = 0.00299244)$. Panel~(ii): $\alpha=0.1$ with critical point $(v_c = 4.92366,\,T_c = 0.0352631,\,P_c = 0.00268574)$. For $T<T_c$ the isotherms exhibit Van der Waals-type oscillations, signalling a small-BH/large-BH first-order transition, while for $T>T_c$ they become monotonic. The shift of the critical point with $\alpha$ reflects the role of the string cloud in controlling the BH phase structure.}
\label{fig:critical}
\end{figure}
The $P$--$v$ isotherms for the LV EKR black hole with a string cloud are shown in Fig.~\ref{fig:critical}. For fixed Lorentz-violating parameter $\ell=-0.1$ and charge $Q=1$, we plot $P$ versus $v$ for two representative values of the string cloud parameter: Fig.~\ref{fig:critical}(i) corresponds to $\alpha=0.05$, while Fig.~\ref{fig:critical}(ii) corresponds to $\alpha=0.1$. In each panel, the three curves correspond to temperatures below, at, and above the critical temperature $T_c$, with the critical isotherm (labelled by $(v_c,T_c,P_c)$ in the caption) separating the regime with Van der Waals–like oscillations ($T<T_c$), where a first-order small-black-hole/large-black-hole phase transition can occur, from the monotonic regime ($T>T_c$), where only a single stable phase exists. For $T<T_c$ the isotherms develop the characteristic oscillatory behavior, signalling the coexistence of small and large black hole phases and the presence of a first-order transition, while for $T=T_c$ the inflection point marks the second-order critical point and for $T>T_c$ the curves become smooth and monotonic. As $\alpha$ increases from Fig.~\ref{fig:critical}(i) to Fig.~\ref{fig:critical}(ii), both $v_c$ and $P_c$ are shifted in agreement with Eq.~\eqref{gg9}, with the critical point moving toward larger specific volumes and lower pressures and temperatures; consequently, the characteristic oscillatory region in the $P$--$v$ diagram is displaced, illustrating how the string cloud parameter tunes the onset of critical behaviour and pushes the small-/large–black-hole transition to larger horizon radii and softer thermodynamic conditions.

In summary, the presence of the Kalb-Ramond LV parameter and the cloud of strings modifies the location of the critical point and the associated scales $(v_c,T_c,P_c,V_c)$, but leaves the universal ratio $P_c v_c/T_c=3/8$ unchanged. This shows that the charged AdS black hole in EKR bumblebee gravity with a string cloud remains in the same thermodynamic universality class as the RN-AdS black hole~\cite{KubiznakMann2012,ZhangCaiYu2015}, while providing an additional handle, through $\ell$ and $\alpha$, to tune the size and temperature of the critical transition.

\section{Thermodynamic Topology}\label{sec:6}

In the previous sections, we identified Van der Waals-like critical behavior from the $P$--$V$ and $T_H(r_+)$ analyzes, but the underlying phase structure can be further clarified from a topological viewpoint. In particular, the method proposed in Refs.~\cite{Wei2022a,Wei2022b} regards black hole configurations as topological defects in an auxiliary parameter space and characterizes their stability in terms of winding numbers. This thermodynamic-topology approach has since been applied to a broad class of AdS and dS black holes in general relativity and modified gravity~\cite{Yerra2022GB,Du2023dS,Sekhmani2025LVKR}, confirming that distinct phase structures can be grouped into a small number of topological classes. In this section, we apply this formalism to the charged AdS black hole in EKR bumblebee gravity with a string cloud, showing how the LV parameter $\ell$ and the string parameter $\alpha$ affect the thermodynamic topology without changing the overall topological class of the solution.

We studied the Hawking temperature of the black hole in the previous section and observed that there is a phase transition for some selected parameters in its curve at $\partial_{r_+}T_{\rm H}\big|_{r_+=r_c}=0$; however, we did not analyze the nature of this phase transition. Here, we refine this analysis by examining the zero points and winding of an appropriate vector field constructed from the generalized free energy, which allows us to distinguish between thermodynamically stable and unstable branches and to identify the type of critical point involved~\cite{Wei2022a,Wei2022b}.

In addition, we analyze the phase transitions present in the generalized free energy of the black hole outside the horizon shell. In doing so, we will also determine the topological class to which the black hole solution belongs. Therefore, we define a potential dependent on the Hawking temperature of the black hole in the form of
\begin{eqnarray}
\Phi =\frac{1}{\sin \theta} T_{\rm H}.\label{ff1}
\end{eqnarray}

Wei {\it et al.}~\cite{Wei2022a,Wei2022b} have introduced another method for studying topological thermodynamics, focusing on generalized free energy functions. This construction introduces an auxiliary two-dimensional space $(r_+,\theta)$, where $\theta$ plays the role of an angular parameter, and the black hole solutions appear as defects (zero points) of a suitably defined vector field. This approach treats black holes as defects in the thermodynamic parameter space, with their solutions explored using the generalized off-shell free energy. In this framework, the stability and instability of black hole solutions are indicated by positive and negative winding numbers, respectively, a picture that has been corroborated in a variety of AdS and dS backgrounds~\cite{Yerra2022GB,Du2023dS,Sekhmani2025LVKR}. 

Given the relationship between mass and energy in black holes, we express the generalized free energy function as a standard thermodynamic function in the following form~\cite{Wei2022a,Wei2022b}
\begin{equation}
\mathcal{F}=M(r_{+})-\frac{S}{\tau},\label{ff3}
\end{equation}
where $\tau$ is the inverse temperature outside the horizon, and $S=\pi r_{+}^2$ represents the entropy of the black hole. The generalized free energy is on-shell only when $\tau=1/T_H$.

The vector space $\boldsymbol{\phi}^{\mathcal{F}}$ of this potential can be represented as~\cite{Wei2022a,Wei2022b}
\begin{align}
\phi^{\mathcal{F}}_{r_{+}}&=\partial_{r_{+}}\mathcal{F}=\frac{1}{2}\left[\frac{1-\alpha}{1-\ell}-\frac{Q^2}{(1-\ell)^2 r^2_{+}}+8 \pi P\,r^2_{+}\right]-\frac{2\pi r_{+}}{\tau},\nonumber\\
\phi_\theta^{\mathcal{F}}&=-\cot \theta \csc \theta.\label{ff5}
\end{align}
normalized according to the following condition: namely
\begin{equation}
    ||\phi||=\sqrt{(\phi^{\mathcal{F}}_{r_{+}})^2+(\phi^{\mathcal{F}}_\theta)^2}.\label{ff2a}
\end{equation}
The normalized vector space ${\bf n}_{\mathcal{F}}$ is therefore given by
\begin{align}
    n^{\mathcal{F}}_{r_{+}}=\frac{\phi^{\mathcal{F}}_{r_{+}}}{||\phi||}\quad,\quad n^{\mathcal{F}}_{\theta}=\frac{\phi^{\mathcal{F}}_{\theta}}{||\phi||}.\label{ff2aa}
\end{align}

The zero points of the vector field $\boldsymbol{\phi}^{\mathcal{F}}$ correspond to $\theta=\pi$ and to the roots of $\partial_{r_{+}}\mathcal{F}=0$, i.e. to stationary points of the generalized free energy in the radial direction. These zero points are in one-to-one correspondence with the thermodynamic phases (e.g. small and large black holes), and their winding numbers determine whether a given phase is locally stable (positive winding) or unstable (negative winding). The zero points of this potential are located at $\theta=\pi$ and $\partial_{r_{+}}\mathcal{F}\big|_{r=r_c}=0$. Moreover, $\phi_\theta^{\mathcal{F}}$ diverges, and the vector direction points outward at $\theta=0$ and $\theta=\pi$. The ranges for $r_{+}$ are $0 \leq r_{+} \leq\infty$ and $\theta$ is $0 \leq \theta \leq \pi$. 

By plotting the unit vector field ${\bf n}_{\mathcal{F}}$ in the $(r_+,\theta)$ plane for fixed values of $(\ell,\alpha,Q,P)$, one can count the total topological charge $W$ enclosed by a large loop. In our model, this total charge is found to be $W=1$, the same value as in the RN--AdS case~\cite{Wei2022a,Wei2022b}, and consistent with the topological classification obtained for other AdS and dS black holes~\cite{Du2023dS,Sekhmani2025LVKR}, indicating that the introduction of the LV parameter and the string cloud does not change the global thermodynamic topology of the black hole, but only deforms the position and nature of the individual zero points.

\begin{figure}[ht!]
    \centering
    \includegraphics[width=0.49\linewidth]{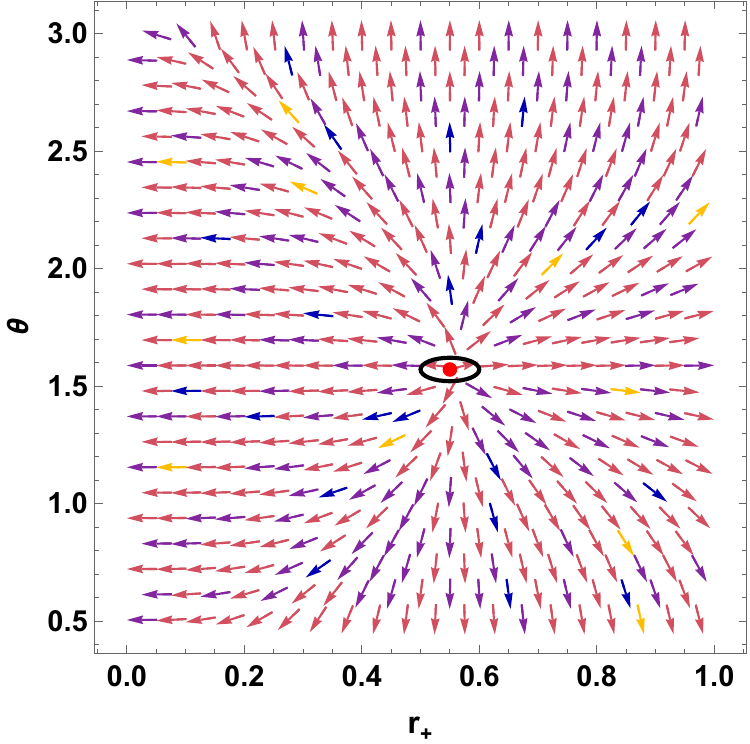}
    \includegraphics[width=0.49\linewidth]{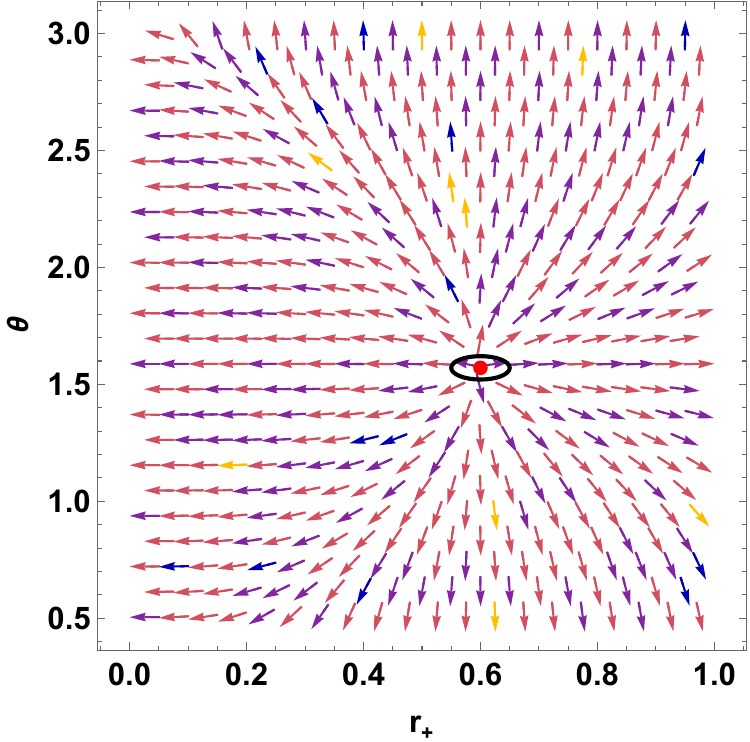}\\
    (i) $\alpha=0.1$ \hspace{2.8cm} (ii) $\alpha=0.2$\\[0.2cm]
    \includegraphics[width=0.49\linewidth]{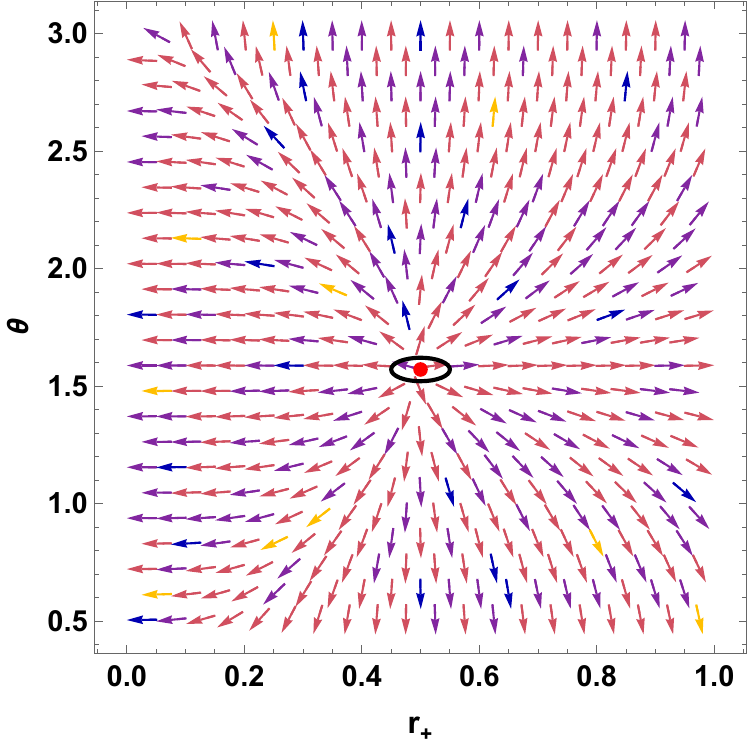}
    \includegraphics[width=0.49\linewidth]{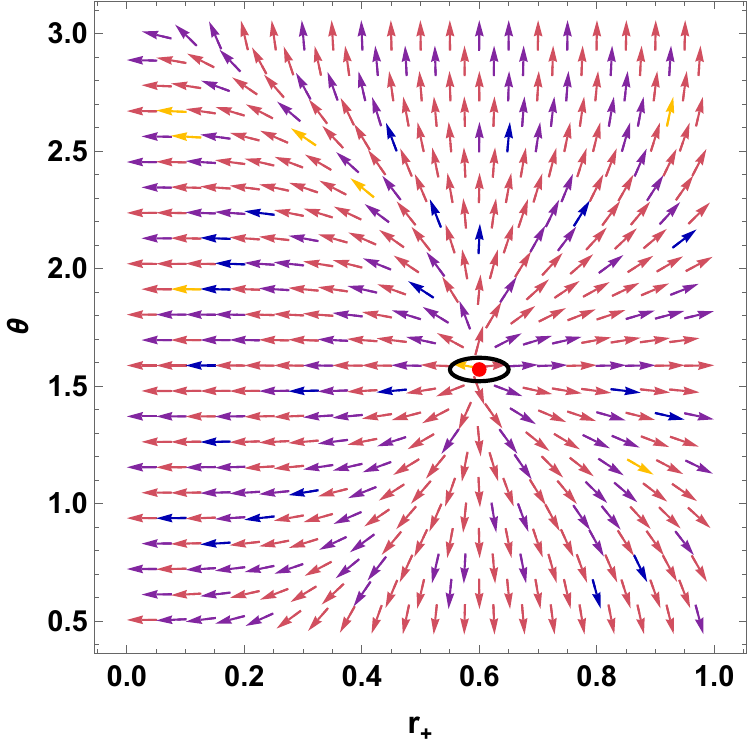}\\
    (iii) $\ell=-0.2$ \hspace{2.6cm} (iv) $\ell=0.2$
    \caption{\footnotesize
    The arrows represent the normalized vector field $\boldsymbol{n}_{\mathcal{F}}$ associated with the generalized free energy on a portion of the $r_{+}$--$\theta$ plane for the charged AdS black hole, obtained by varying the parameters $\alpha$ and $\ell$. Panels~(i) and~(ii) display the vector field for fixed $\ell=0.1$ and two values of the string cloud parameter, $\alpha=0.1$ and $\alpha=0.2$, respectively; in these cases, the contour surrounding the red bullet corresponds to the point $(r_{+},\theta) = (r_{+},\pi/2)$ with $r_{+}=0.55$ in panel~(i) and $r_{+}=0.6$ in panel~(ii). Panels~(iii) and~(iv) show the effect of varying the Lorentz-violating parameter, $\ell=-0.2$ and $\ell=0.2$, respectively, at fixed $\alpha=0.1$, with $r_{+}=0.5$ in panel~(iii) and $r_{+}=0.6$ in panel~(iv). In all plots we use the parameter values $Q = 0.5$, $P = \tfrac{0.03}{8\pi}$, and $\tau = 20\pi$.}
    \label{fig:unit-vector}
\end{figure}
The structure of the corresponding normalized vector field $\boldsymbol{n}_{\mathcal{F}}$ in the auxiliary $(r_{+},\theta)$ space is displayed in Fig.~\ref{fig:unit-vector}. For fixed $(Q,P,\tau)$, each panel shows how the arrows of $\boldsymbol{n}_{\mathcal{F}}$ wind around a single zero of the vector field (indicated by the red point), which corresponds to a stationary point of the generalized free energy $\mathcal{F}(r_{+},\tau)$. Figures~\ref{fig:unit-vector}(i) and \ref{fig:unit-vector}(ii) illustrate the effect of increasing the string cloud parameter $\alpha$ at fixed $\ell=0.1$, while Figs. \ref{fig:unit-vector}(iii) and \ref{fig:unit-vector}(iv) show the effect of varying the Lorentz-violating parameter $\ell$ at fixed $\alpha=0.1$. In all cases, changing $\alpha$ or $\ell$ displaces the location of the zero in the $(r_{+},\theta)$ plane but does not change its winding, so that the total topological charge enclosed by a large contour remains $W=1$. This confirms that the EKR black hole with a cloud of strings belongs to the same thermodynamic topological class as the RN–AdS solution, and that $\ell$ and $\alpha$ act only as deformation parameters that shift the positions of the stable and unstable branches without altering the global topology of the phase space.

\begin{figure}[ht!]
    \centering
    \includegraphics[width=0.75\linewidth]{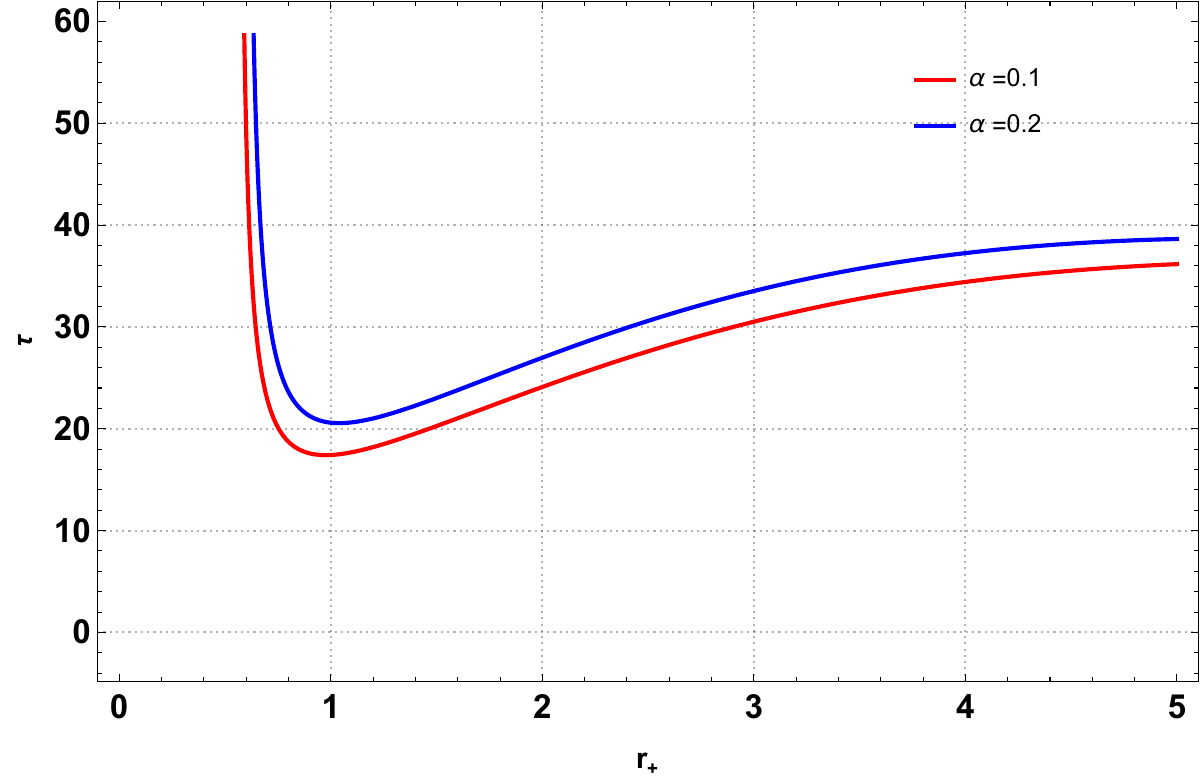}\\
    (i) $\ell=0.1$\\
    \includegraphics[width=0.75\linewidth]{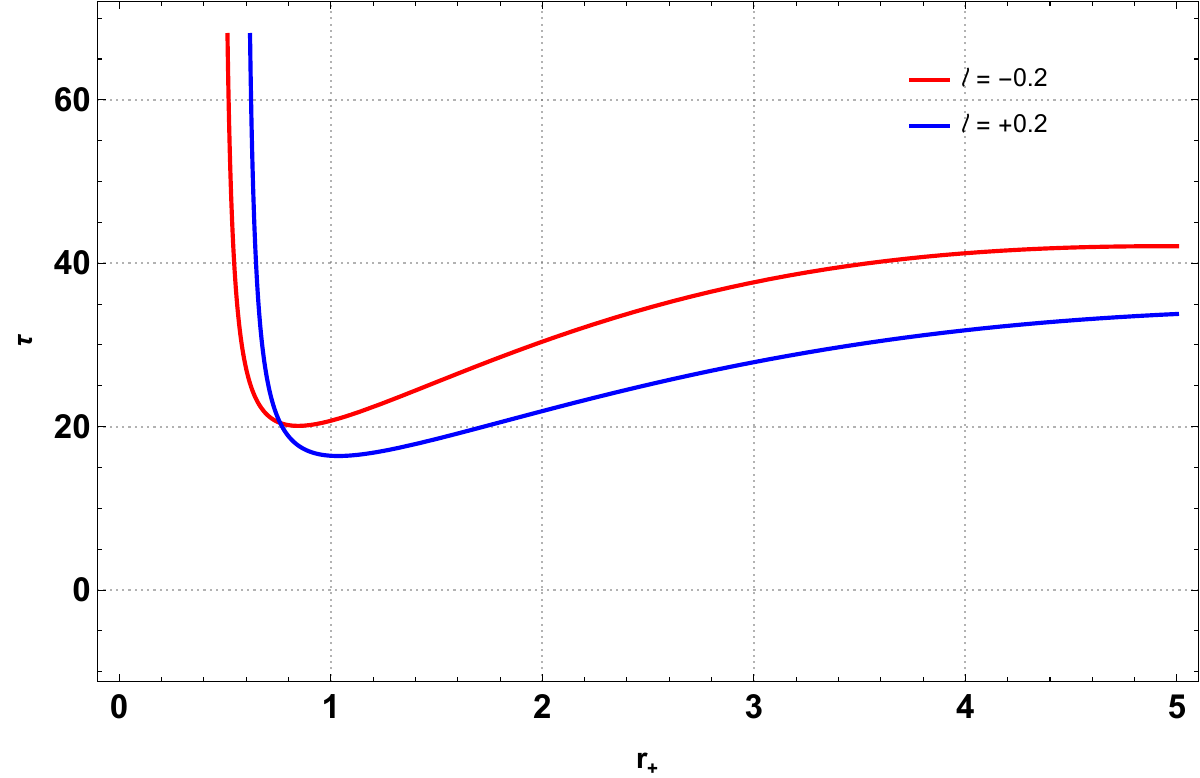}\\
    (ii) $\alpha=0.1$
    \caption{\footnotesize Behavior of the Euclidean time $\tau$ (inversion temperature) as a function of horizon for the charged AdS BH. Here $Q=0.5,\,P=\tfrac{0.03}{8\pi}$.}
    \label{fig:inverse-temperature}
\end{figure}

Next, we find the zero points of the $\phi^{\mathcal{F}}_{r_{+}}$ component by solving $\phi^{\mathcal{F}}_{r_{+}}=0$ and derive an expression for the Euclidean time period $\tau$. In terms of horizon $r_{+}$, we find 
\begin{equation}
\tau=\frac{2\pi r_{+}}{M'(r_{+})}=\frac{4\pi r_{+}}{\frac{1 - \alpha}{1 - \ell} - \frac{Q^2}{(1 - \ell)^2 r_+^2} +8\pi P\, r_+^2}=1/T_H.\label{ff6}
\end{equation}
This confirms that the zero points of $\boldsymbol{\phi}^{\mathcal{F}}$ coincide with the on-shell configurations $T_H=1/\tau$, so that the topological classification is directly tied to the physical Hawking temperature and the phase structure found in the previous section.

\section{Joule--Thomson expansion: Inversion temperature}

In this section we study the JT-expansion of charged AdS black hole with string cloud within the bumblebee gravity. In the extended phase space, where the black hole mass plays the role of enthalpy, the Joule--Thomson process is naturally represented by isenthalpic curves $M=\mathrm{const.}$ in the $T$--$P$ plane. For brevity, we will still refer to these as ``isenthalpic'' curves, even though, strictly speaking, they are constant-mass trajectories in the black hole parameter space. Joule--Thomson expansion in AdS black hole backgrounds has been extensively investigated for Reissner--Nordström--AdS, Kerr--AdS, Gauss--Bonnet, Lovelock and other deformed black holes~\cite{Okcu2017,Mo2018JT,Lan2018JT,MoLi2020Lovelock,Cisterna2019JT,Rizwan2019Monopole,Chabab2018fR}, providing a useful benchmark for our Lorentz-violating setup with a string cloud.

Recall the expression for Joule Thomson coefficient
 \begin{equation}
  \mu_J=\frac{\partial T}{\partial P}=\frac{1}{C_{\rm heat}}\left[T \left(\frac{\partial V}{\partial T}\right)_P-V\right].\label{pp1}   
 \end{equation}
From this we obtain the inversion temperature
\begin{equation}
    T_i=V \left(\frac{\partial T}{\partial V}\right)_P.\label{pp2} 
\end{equation} 

\begin{figure}[ht!]
    \centering
    \includegraphics[width=0.85\linewidth]{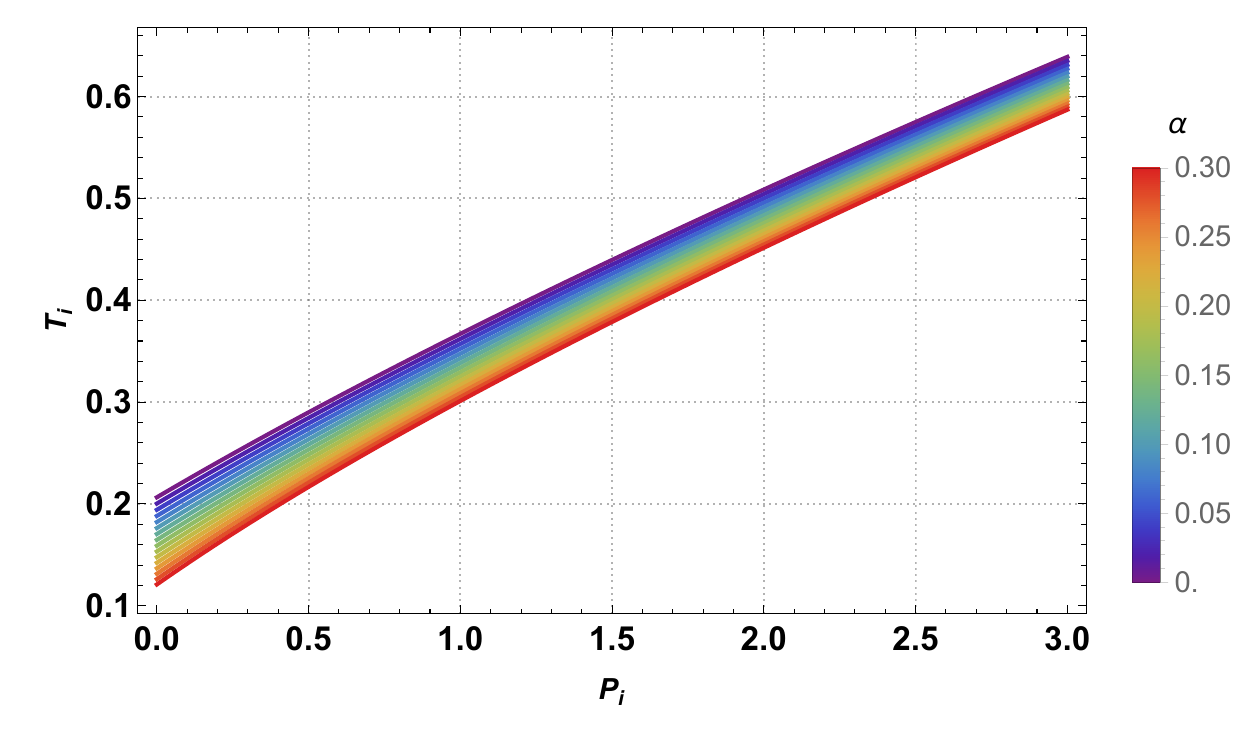}\\
    (i) $\ell=-0.1$\\
    \includegraphics[width=0.85\linewidth]{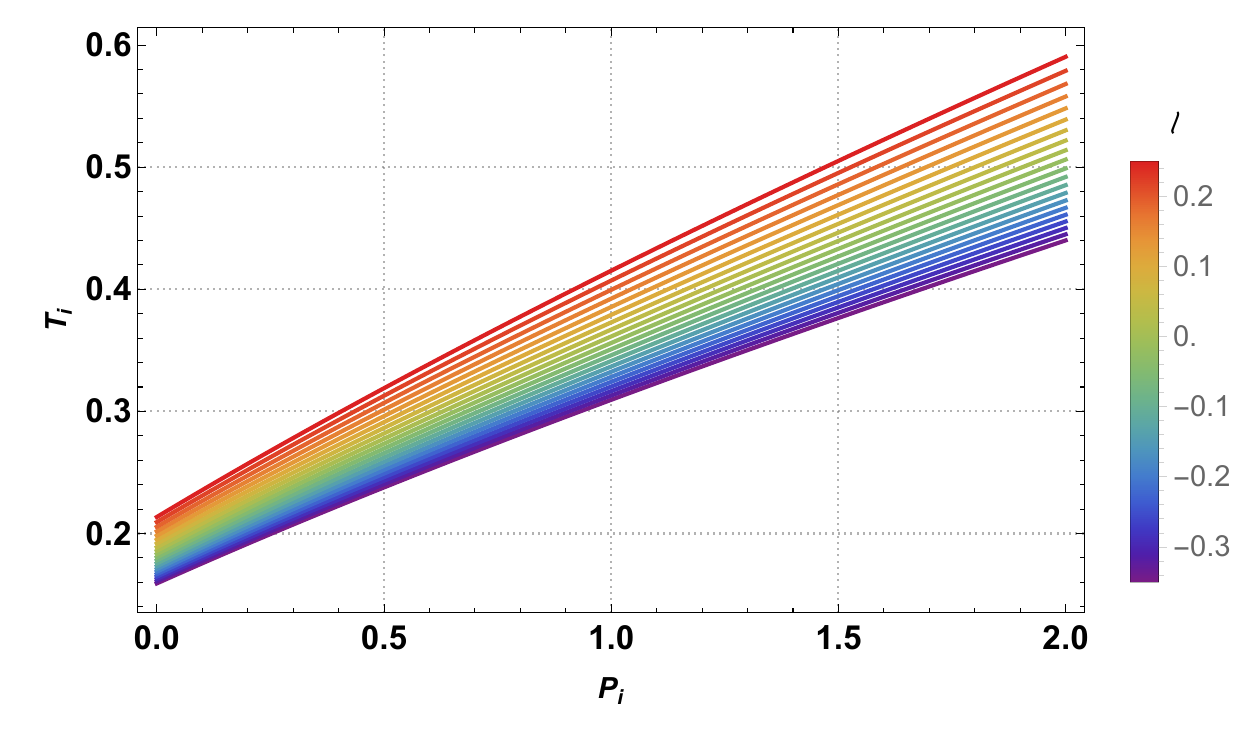}\\
    (ii) $\alpha=0.1$
    \caption{\footnotesize Effect of string cloud parameter $\alpha$ and KR-field parameter $\ell$ on inversion curve. Here we have chosen different values of $\alpha$ and $\ell$ by keeping charge $Q=0.1$ fixed.}
    \label{fig:inversion-curve}
\end{figure}

\begin{figure}[ht!]
    \centering
    \includegraphics[width=0.75\linewidth]{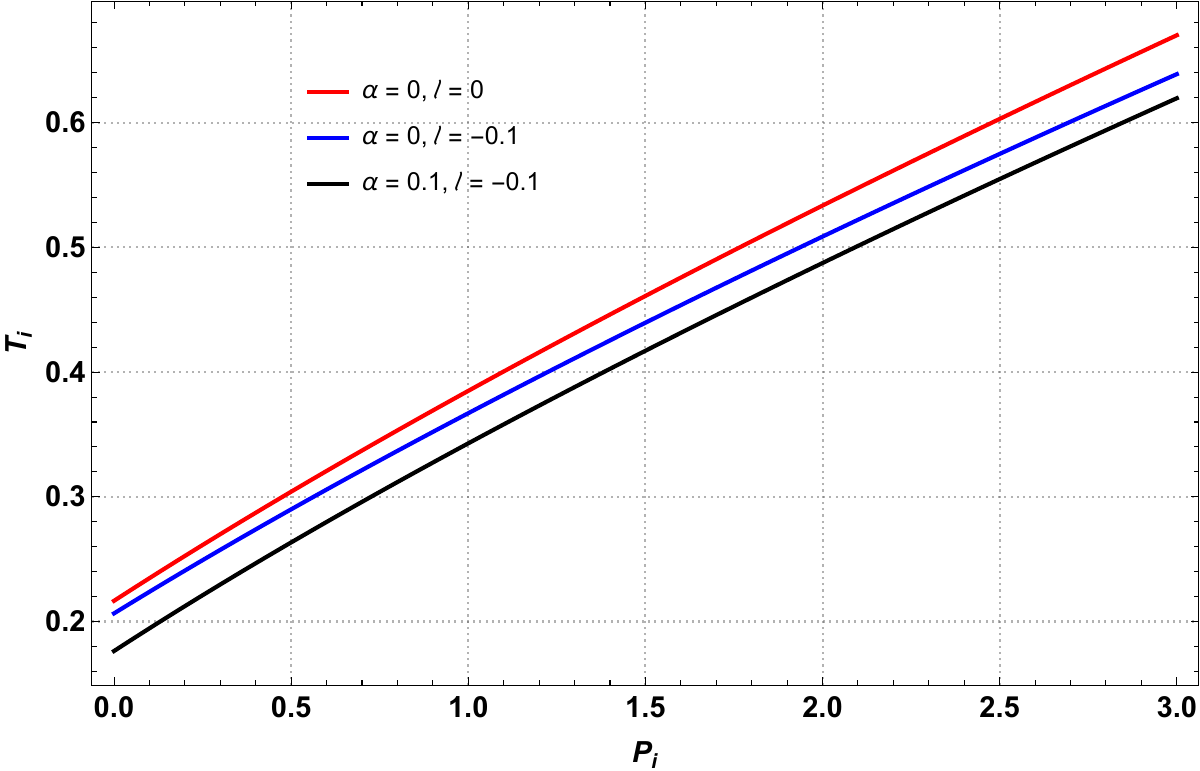}
    \caption{\footnotesize Inversion curve for charged AdS BH with the combinations of string cloud and KR-field parameters by keeping $Q=01$ fixed. The plots are the locus of inversion points ($P_i, T_i$).}
    \label{fig:inversion-curve-comparison}
\end{figure}

The temperature $T=T_H$ in terms of volume $V$ from Eq. (\ref{gg4}) can be expressed as
\begin{align}
    T=\frac{1}{4\pi}\frac{1 - \alpha}{1 - \ell}\left(\frac{4\pi}{3}\right)^{1/3}\,&V^{-1/3}-\frac{Q^2}{3 (1 - \ell)^2 V}\notag\\&+2 P \left(\frac{3}{4\pi}\right)^{1/3} V^{1/3}.\label{pp3}
\end{align}
Therefore, we find the inversion temperature in terms of horizon 
\begin{align}
    T_i&=-\frac{1}{12\pi r_{+}}\frac{1 - \alpha}{1 - \ell}+\frac{Q^2}{4\pi (1 - \ell)^2 r^3_{+}}+\frac{2 P}{3} r_{+}.\label{pp4}
\end{align}
Using Eqs. (\ref{gg4}) and (\ref{pp4}), we find the following relation for $T=T_i$ as
\begin{equation}
    \frac{1 - \alpha}{1 - \ell} r^2_{+}-\frac{3 Q^2}{2(1 - \ell)^2}+4 \pi P r^4_{+}=0.\label{pp5}
\end{equation}
 Solving the above for $r_{+}$ and choosing the following appropriate root,
 \begin{equation}
     r_{+}=\sqrt{\frac{-(1-\alpha) + \sqrt{(1-\alpha)^{2} + 24\pi P Q^{2}}}{8\pi P\, (1-\ell)}}\label{pp6}
 \end{equation}
Substituting this root into the Eq. (\ref{pp4}) yields
\begin{align}
T_i&=\frac{1}{4\pi}
\sqrt{\frac{8\pi P_i (1-\ell)}{-(1-\alpha)+\sqrt{(1-\alpha)^2+24\pi P_i Q^{2}}}}
\Bigg[
\frac{1-\alpha}{1-\ell}\nonumber\\
&-\frac{8\pi P_i Q^{2}}{(1-\ell)\!\left(-(1-\alpha)+\sqrt{(1-\alpha)^2+24\pi P_i Q^{2}}\right)}\nonumber\\
&+\frac{-(1-\alpha)+\sqrt{(1-\alpha)^2+24\pi P_i Q^{2}}}{1-\ell}
\Bigg].\label{pp7}
\end{align}
From the above equation (\ref{pp7}), we see that the inversion temperature is influenced by both KR-field parameter $\ell$ and string cloud parameter $\alpha$ including the electric charge $Q$. From this equation, the inversion curves are plotted for different values of $\alpha$ and $\ell$. Figure \ref{fig:inversion-curve}, we study the inversion curves for various values of $\alpha$ and $\ell$ in the $T$-$P$ plane. Figure \ref{fig:inversion-curve-comparison} shows an inversion curve with and without string cloud and KR-field parameters.

By demanding $P_i = 0$, we obtain 
\begin{equation}
    T^{\rm min}_i=\frac{(1-\alpha)^{3/2}}{12\pi Q}
\sqrt{\frac{2}{3(1-\ell)}}.\label{pp8}
\end{equation}
Using this we calculate the ratio between $T^{\rm min}_i$ and $T_c$ as, 
\begin{equation}
    \frac{T^{\rm min}_i}{T_c}=\frac{1}{2}.\label{pp9}
\end{equation}
This result shows that, despite the presence of the LV parameter and the string cloud, the minimal inversion temperature remains exactly one half of the critical temperature,
in agreement with the behaviour of charged AdS black holes in Einstein gravity~\cite{Okcu2017}. The isenthalpic curves in the $T$--$P$ plane therefore split into cooling and heating regions across the inversion curve $T_i(P)$, and the ratio $T^{\rm min}_i/T_c=1/2$ confirms that the Joule--Thomson expansion of the EKR bumblebee black hole with string cloud belongs to the same universality class as the RN--AdS case, while providing new control parameters $(\ell,\alpha)$ to tune the absolute scales of $T_i$ and $T_c$.

Beyond the location of the critical point, it is also useful to determine the coexistence line separating small- and large-black-hole phases in the $P$--$T_H$ plane. For $T_H<T_c$, the Van der Waals-like oscillations in the isotherms of Fig.~\ref{fig:critical} signal a first-order phase transition that can be treated using the Maxwell equal-area construction. Introducing the specific volumes $v_s$ and $v_l$ of the small and large black holes at a given temperature $T_H<T_c$, the equal-area law reads
\begin{equation}
\int_{v_s}^{v_l} P(T_H,v)\,\dd v
= P_{\rm tr}(T_H)\,\bigl(v_l - v_s\bigr),
\label{eq:Maxwell}
\end{equation}
where $P_{\rm tr}(T_H)$ is the transition pressure at that temperature. Together with the conditions
\begin{equation}
P\bigl(T_H,v_s\bigr)=P\bigl(T_H,v_l\bigr)=P_{\rm tr}(T_H),
\end{equation}
Eq.~\eqref{eq:Maxwell} determines the coexistence curve $P_{\rm tr}(T_H)$ implicitly. Since our equation of state~\eqref{gg6} has the same functional form as that of the RN--AdS black hole, the structure of the coexistence line is identical to the Van der Waals case~\cite{KubiznakMann2012,ZhangCaiYu2015}, while the parameters $\ell$ and $\alpha$ enter through the effective couplings $\eta(\ell,\alpha)$ and $\zeta(\ell,Q)$, shifting the coexistence region in the $(P,T_H)$ plane. In practice, the system of equations for $(v_s,v_l,P_{\rm tr})$ can be solved numerically for fixed $(\ell,\alpha,Q)$, allowing one to map out how Lorentz violation and the string cloud deform the small-/large-black-hole coexistence line.

The slope of the coexistence line follows from the usual Clapeyron relation,
\begin{equation}
\frac{\dd P_{\rm tr}}{\dd T_H}
= \frac{\Delta S}{\Delta V}
= \frac{\pi\,(r_{+,l}^2-r_{+,s}^2)}
{\tfrac{4\pi}{3}\,(r_{+,l}^3-r_{+,s}^3)},
\label{eq:Clapeyron}
\end{equation}
where $(r_{+,s},r_{+,l})$ are the horizon radii of the coexisting small and large black holes, and $\Delta S$ and $\Delta V$ are the entropy and volume jumps across the transition. Equation~\eqref{eq:Clapeyron} makes explicit that the latent heat and the slope of the coexistence curve are entirely governed by the geometric data $(r_{+,s},r_{+,l})$, while the Lorentz-violating and string-cloud parameters affect them indirectly by shifting the radii at which coexistence occurs.

To further quantify the critical behaviour, we now extract the critical exponents associated with the equation of state~\eqref{gg6}. Introducing reduced variables,
\begin{equation}
p=\frac{P}{P_c},\qquad
\tau=\frac{T_H}{T_c},\qquad
\nu=\frac{v}{v_c},
\end{equation}
and expanding $p(\tau,\nu)$ near the critical point $(p,\tau,\nu)=(1,1,1)$ by setting $\tau=1+t$ and $\nu=1+\omega$ with $|t|\ll 1$ and $|\omega|\ll 1$, one finds the generic Landau-type expansion
\begin{equation}
p = 1 + A\,t - B\,t\,\omega - C\,\omega^3 + \mathcal{O}\!\left(t\,\omega^2,\omega^4\right),
\label{eq:EoS-expand}
\end{equation}
where $A,B,C>0$ are constants that depend on $(\ell,\alpha,Q)$ through $\eta$ and $\zeta$, but do not affect the scaling structure. Equation~\eqref{eq:EoS-expand} has the same form as in the RN--AdS and Van der Waals cases~\cite{KubiznakMann2012,ZhangCaiYu2015}, and therefore yields the standard mean-field critical exponents.

Following the usual procedure, one obtains $\alpha_0=0$ from the regularity of the specific heat at constant volume, $C_V\propto |t|^{-\alpha_0}$; the discontinuity of the order parameter $\eta_v\equiv v_l-v_s\propto |t|^{\beta_0}$ along the coexistence line implies $\beta_0=\tfrac{1}{2}$; the isothermal compressibility $\kappa_T\propto |t|^{-\gamma_0}$ diverges with $\gamma_0=1$; and, finally, the critical isotherm $t=0$ satisfies $|p-1|\propto |\omega|^{\delta_0}$ with $\delta_0=3$. We thus find
\begin{equation}
\alpha_0=0,\qquad
\beta_0=\frac{1}{2},\qquad
\gamma_0=1,\qquad
\delta_0=3,
\end{equation}
showing that the charged AdS black hole in EKR-gravity with a cloud of strings belongs to the same mean-field universality class as the RN-AdS black hole and the Van der Waals fluid, despite the nontrivial dependence of the critical scales $(v_c,T_c,P_c)$ on the Lorentz-violating parameter $\ell$ and the string cloud parameter $\alpha$.

\section{Summary and conclusions}\label{sec4}

In this work we have investigated the thermodynamic properties and phase structure of electrically charged AdS black holes in Einstein--Kalb--Ramond (EKR) gravity in the presence of a spherically symmetric cloud of strings. Starting from the static, spherically symmetric solution supported by the background Kalb--Ramond (KR) field and characterized by the Lorentz-violating parameter $\ell$~\cite{ref1,ref2} and by the string cloud parameter $\alpha$~\cite{PSL}, we constructed a charged AdS black hole metric whose lapse function $f(r)$ encodes both Lorentz symmetry violation and the matter distribution associated with the string cloud. In the limit $\alpha\to 0$ our solution reduces to the charged AdS black hole in EKR gravity of Ref.~\cite{ref2}, while for $\ell\to 0$ it reproduces the standard charged AdS black hole with a cloud of strings in Einstein gravity, thereby interpolating smoothly between previously known configurations.

Within the extended phase space formalism, where the cosmological constant is interpreted as a thermodynamic pressure and the ADM mass is identified with the enthalpy of the system~\cite{PhysRevD.102.044028,KubiznakMann2012}, we derived explicit expressions for the thermodynamic variables of the EKR black hole with string cloud. In particular, we obtained the ADM mass $M(r_+)$, the thermodynamic volume $V=\tfrac{4\pi}{3}r_+^3$, the Hawking temperature $T_H$, the Gibbs free energy $F$, the internal energy $U$, and the specific heat $C_{\rm heat}$ as functions of the horizon radius and of the parameters $(\ell,\alpha,Q,P)$. The Lorentz-violating parameter and the string cloud both act as effective ``deformation'' parameters of the Reissner--Nordström--AdS thermodynamics: they shift the zeros and extrema of $T_H(r_+)$, modify the structure of the Gibbs free energy, and change the sign and divergence points of $C_{\rm heat}$, thereby controlling the ranges of local thermodynamic stability, in close analogy with other extended-phase-space black hole models~\cite{KubiznakMann2012,Okcu2017}. The qualitative behavior of these quantities is illustrated by our plots of $M(r_+)$, $T_H(r_+)$, $F(r_+)$, $U(r_+)$ and $C_{\rm heat}(r_+)$, which show how increasing $\alpha$ or varying $\ell$ effectively ``dresses'' the black hole for a given horizon radius.

We then analyzed the $P$--$V$ criticality of the model by rewriting the equation of state in the Van der Waals form $P(T_H,v)$, with $v=2r_+$. Solving the criticality conditions $(\partial P/\partial v)_{T_H}=0$ and $(\partial^2 P/\partial v^2)_{T_H}=0$, we obtained analytic expressions for the critical specific volume $v_c$, temperature $T_c$ and pressure $P_c$. A notable result is that $P_c$ depends only on the string cloud parameter $\alpha$, whereas both $v_c$ and $T_c$ are sensitive to the Lorentz-violating parameter $\ell$ as well as to $\alpha$. Nevertheless, the universal ratio $P_c v_c/T_c=3/8$ is preserved, exactly as in the Reissner--Nordström--AdS case~\cite{KubiznakMann2012}, showing that the EKR black hole with a cloud of strings lies in the same mean-field universality class as ordinary Van der Waals fluids. The numerical values presented in Tables~\ref{tab:1}--\ref{tab:3} quantify how the Lorentz-violating background and the string cloud deform the critical size, temperature and pressure relative to the RN--AdS benchmark. Consistently with this analysis, the $P$--$v$ isotherms in Fig.~\ref{fig:critical} exhibit Van der Waals–like oscillations for $T_H<T_c$ and a second-order critical point at $(v_c,T_c,P_c)$, with the oscillatory region and the location of the critical point shifted as $\alpha$ is varied.

To further characterize the phase structure, we employed the thermodynamic topology approach of Wei \textit{et al.}~\cite{Wei2022a,Wei2022b}, constructing a generalized off-shell free energy $\mathcal{F}(r_+,\tau)$ and the associated vector field in the auxiliary $(r_+,\theta)$ space. The zero points of this field correspond to black hole phases, and their winding numbers distinguish stable from unstable branches. Our analysis shows that the total topological charge of the EKR black hole with string cloud is $W=1$, identical to the RN--AdS case~\cite{Wei2022a,Wei2022b} and consistent with other AdS and dS black holes~\cite{Du2023dS,Sekhmani2025LVKR}. As illustrated in Fig.~\ref{fig:unit-vector}, changing the parameters $\ell$ and $\alpha$ displaces the zero of the normalized vector field in the $(r_+,\theta)$ plane but does not alter its winding. This indicates that the LV parameter and the string cloud do not change the global thermodynamic topology of the solution, but only shift the positions and parameter ranges in which multiple phases coexist.

Finally, we studied the Joule--Thomson expansion of the black hole by computing the inversion temperature $T_i(P)$ from the Joule--Thomson coefficient and the equation of state. We derived an analytic expression for the inversion curve and the minimal inversion temperature $T_i^{\rm min}$, and showed that the ratio $T_i^{\rm min}/T_c=1/2$ is exactly the same as for charged AdS black holes in Einstein gravity~\cite{Okcu2017,Mo2018JT,Lan2018JT,MoLi2020Lovelock,Cisterna2019JT,Rizwan2019Monopole,Chabab2018fR}. Thus, even in the presence of Lorentz violation and a cloud of strings, the qualitative structure of the isenthalpic curves in the $T$--$P$ plane and the division between heating and cooling regions remain universal, while the absolute scales of $T_i$ and $T_c$ can be tuned by the parameters $(\ell,\alpha)$.

Overall, our results indicate that the combination of a background KR field and a cloud of strings provides a flexible framework in which Lorentz-violating effects can be incorporated into black hole thermodynamics without spoiling the universal critical behaviour. The parameters $(\ell,\alpha)$ act as geometric and matter ``knobs'' that deform the thermodynamic scales and stability windows, but leave the Van der Waals universality class, the topological charge, and the Joule--Thomson ratios unchanged. It would be natural, in future work, to extend this analysis to rotating solutions and higher-dimensional generalizations, to investigate dynamical stability and quasinormal modes in the same background, and to explore possible holographic interpretations of the Lorentz-violating and string-cloud sectors in dual field theories.

\small 
\section*{Acknowledgments}

F.A. acknowledges the Inter University Centre for Astronomy and Astrophysics (IUCAA), Pune, India for granting visiting associateship. E. O. Silva acknowledges the support from grants CNPq/306308/2022-3, FAPEMA/UNIVERSAL-06395/22, FAPEMA/APP-12256/22, and (CAPES) - Brazil (Code 001).

\section*{Data Availability Statement}

This manuscript has no associated data. [Authors’ comment: Data sharing not applicable to this article as no datasets were generated or analyzed during this study.]

\section*{Code Availability Statement}

This manuscript has no associated code/software. [Authors’ comment: Code/Software sharing not applicable to this article as no code/software was generated or analysed during this study.]

\bibliographystyle{apsrev4-2}
%\bibliography{References}
\input{Article.bbl}

\end{document}

%% file: Article.bbl
%apsrev4-2.bst 2019-01-14 (MD) hand-edited version of apsrev4-1.bst
%Control: key (0)
%Control: author (72) initials jnrlst
%Control: editor formatted (1) identically to author
%Control: production of article title (-1) disabled
%Control: page (0) single
%Control: year (1) truncated
%Control: production of eprint (0) enabled
%